%% file: main.tex
\documentclass[twocolumn, superscriptaddress]{revtex4-2}
\usepackage{graphicx} 
\usepackage[caption=false]{subfig}
\usepackage{amsmath}
\usepackage{amssymb}
\usepackage{physics}
\usepackage{quantikz}
\usepackage{adjustbox}
\usepackage{hyperref}
\usepackage[nameinlink,capitalise]{cleveref}

\makeatletter
\AddToHook{cmd/appendix/before}{\def\cref@section@alias{appendix}\def\cref@subsection@alias{appendix}}
\makeatother

\begin{document}

\title{Low-valency scalable quantum error correction with a dynamic compass code}

\author{Jun Zen}
\affiliation{Okinawa Institute of Science and Technology, Okinawa, Japan}

\author{Xanda C. Kolesnikow}
\affiliation{School of Physics, The University of Sydney, Sydney, New South Wales 2006, Australia}

\author{Campbell K. McLauchlan}
\affiliation{School of Physics, The University of Sydney, Sydney, New South Wales 2006, Australia}

\author{Georgia M. Nixon}
\affiliation{School of Physics, The University of Sydney, Sydney, New South Wales 2006, Australia}

\author{Thomas R. Scruby}
\affiliation{Okinawa Institute of Science and Technology, Okinawa, Japan}

\author{Seok-Hyung Lee}
\affiliation{School of Physics, The University of Sydney, Sydney, New South Wales 2006, Australia}
\affiliation{Department of Quantum Information Engineering, Sungkyunkwan University, Suwon 16419, Republic of Korea}

\author{Stephen D. Bartlett}
\affiliation{School of Physics, The University of Sydney, Sydney, New South Wales 2006, Australia}

\author{Benjamin J. Brown}
\affiliation{IBM Quantum, T. J. Watson Research Center, Yorktown Heights, New York 10598, USA}
\affiliation{IBM Denmark, Sundkrogsgade 11, 2100 Copenhagen, Denmark}

\author{Robin Harper}
\affiliation{School of Physics, The University of Sydney, Sydney, New South Wales 2006, Australia}

\begin{abstract}
The ongoing development of hardware that is capable of reliably executing general quantum algorithms requires quantum error-correcting codes that are both practical for realisation and rapidly reduce logical error rates as they are scaled up.
Here we introduce the dynamic compass code, a code that can be implemented with a modest footprint on the heavy-hex lattice while also demonstrating a threshold. The dynamic code is obtained by choosing a novel measurement schedule for the syndrome extraction circuit of the heavy-hex subsystem code. 
We numerically evaluate its performance and observe that different choices of schedule can provide a trade-off in protection against logical errors in the $X$ vs $Z$ basis. We also demonstrate that this new measurement schedule provides the code with a threshold for stability experiments. 
We finally show how the dynamic compass code could be used for fault-tolerant logic by illustrating lattice surgery between code patches.
\end{abstract}

\maketitle

\section{Introduction}

Quantum error correction enables reliable quantum computing using a noisy device. To reach a stage where we can perform high-fidelity logic gates on near-ideal qubits in the future we require quantum error-correcting codes with logical error rates that vanish rapidly as they are scaled up, i.e., we need codes that demonstrate a threshold. Ideally these codes will also respect constraints that enable their realisation with modern hardware. To this end two-dimensional planar layout and low-degree connectivity are also desirable. 
The discovery of families of codes that respect these desiderata give us new paths forward to realise a large-scale quantum computer. Here, we present a family of codes that can be realised on a planar lattice with heavy-hex connectivity, i.e. a lattice design where qubits can couple with at most three of their neighbours. Our code construction combines two recent ideas from quantum error-correction theory: compass codes~\cite{Bacon2006operator, Li2019, campos2026clifforddeformedcompasscodes} and dynamic (Floquet) codes~\cite{Hastings2021}.

Compass codes~\cite{Bacon2006operator, Li2019, campos2026clifforddeformedcompasscodes} are obtained by selecting a gauge group that is a subset of the Pauli operators generated by the interaction terms of a compass model~\cite{Kugel1982, KITAEV2006, Nussinov2015compass}.
This framework for code design gives us freedom to control key code parameters that factor into their practical realisation. These parameters include, for instance, the weights of stabilisers and their relative sensitivity to different noise channels.
Well-known examples derived from the square-lattice compass model~\cite{Kugel1982} include the surface code~\cite{Kitaev2003}, the Bacon-Shor code~\cite{Bacon2006operator} and the heavy-hex code~\cite{Chamberland2020topological}. 
Importantly, the compass code framework allows us to interpolate between the code parameters of extremal examples of compass codes by selecting different choices of gauge group elements.

A second innovation we exploit is the notion of dynamic codes. It is possible to design code families with practical quantum readout circuits by carefully considering the order in which the gauge terms are measured, i.e., the circuit dynamics. The primary example is the Hastings-Haah Floquet code~\cite{Hastings2021}; a dynamic code that shares the properties of the surface code. However, the Floquet code detects errors using only weight-two measurements by choosing a specific measurement ordering of the terms of Kitaev's honeycomb model, i.e., the hexagonal-lattice compass model~\cite{Kugel1982, KITAEV2006, Nussinov2015compass}. A great number of dynamic code families have subsequently been proposed~\cite{Gidney2023pairmeasurement, Davydova2023floquet, kesselring2024, Davydova2024quantumcomputation, Dua2024, Higgott2024, Setiawan2025, derks2025dynamicalcodeshardwarenoisy}. Examples of particular relevance here are dynamic codes derived from the square-lattice compass model~\cite{gidney2023baconthreshold, Alam2025dynamical} where a dynamic measurement sequence for the checks of the Bacon-Shor code give rise to a code family that demonstrates a threshold. Ref.~\cite{alam2026baconshorboardgames} explores this idea further to find a dynamic code based on the square-lattice compass model whose detectors occupy a constant volume in spacetime.

Deriving code families from the weight-two interaction terms of compass models necessarily gives rise to practical constructions that require only low-weight checks to read out stabilisers. Furthermore, it has been shown that a dynamic readout schedule can qualitatively change the properties of a code family to enhance their performance~\cite{Hastings2021, gidney2023baconthreshold, alam2026baconshorboardgames}. 
Indeed, we put forward that exploring this approach further could give rise to a general framework to produce other families of dynamic codes from compass models. 
To exemplify this as a design principle to produce families of practical codes, here we define a dynamic compass code by proposing a non-trivial readout schedule for the gauge group of a specific compass code, namely the heavy-hex code~\cite{Chamberland2020topological}.

Unlike its subsystem code precursor, our dynamic compass code demonstrates a threshold for logical errors in both bases.
Additionally, by beginning with the heavy-hex code gauge group, that can be read out with only weight-two and weight-four parity checks, we obtain a dynamic code with a readout circuit that is readily executed on a heavy-hex lattice.
Our novel choice of measurement schedule gives rise to a dynamically evolving stabilizer group with only constant-weight stabilisers.
In contrast, the heavy-hex code with a standard schedule gives rise to some extensive weight stabilisers.
We note that IBM routinely produce large instances of superconducting devices with this connectivity~\cite{sundaresan2022, Gupta2024, harper2025characterising} and that, in an accompanying paper, we implement this code on an IBM r3 Heron class device; Pittsburgh~\cite{DCCexperiment}.

We investigate the family of dynamic compass codes with numerical simulations. Under a circuit-level noise model we investigate a tradeoff in threshold between Pauli-$X$ basis errors and Pauli-$Z$ basis errors with variations in our measurement schedule. In addition to this we study a dynamic compass code through the lens of a stability experiment, thereby demonstrating the code has a vanishing logical error rate for measurement-based logical operations. We also make a comprehensive study of the threshold rate as we vary the relative noise rates of different types of circuit elements, to determine a regime where the dynamic compass code is scalable. We note that with an appropriate schedule the dynamic compass code family shares the essential properties of the surface code. We exploit this property to sketch out how we might scale this model to perform the logic gates needed for fault-tolerant quantum computation via lattice surgery.

The rest of this paper is structured as follows. In \cref{section:prelim} we introduce the heavy-hex code, dynamic compass codes, and discuss detecting regions and physically motivated syndrome extraction circuits for these codes. In \cref{section:decoding} we describe the details of the noise models and decoding approach used in our simulations and then present the results of these simulations in \cref{section:results}. In \cref{section:surgery} we show how to generalize lattice surgery techniques to these codes and then conclude with a general discussion in \cref{section:discussion}.

\section{Preliminaries}
\label{section:prelim}

\begin{figure}
    \centering
    \begin{tikzpicture}
        \node at (-1,0) {\includegraphics[width=0.47\linewidth]{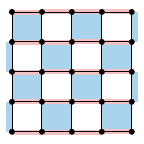}}; 
        \node at (3.4,0) {\includegraphics[width=0.51\linewidth]{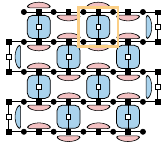}};
        \node[inner sep=0pt] at (-3,2)         {(a)};
        \node[inner sep=0pt] at (1.4,2)         {(b)};
    \end{tikzpicture}
    \caption{The gauge group we study in this work defined in (a) the compass model and (b) on the heavy-hex lattice. In both figures blue regions show weight-4 $X$ measurements, red regions show weight-$2$ $Z$ measurements and circles show data qubits. In (b), empty and filled squares show $X$ and $Z$ measurement qubits respectively and edges show device connectivity. The region in the yellow box corresponds to \cref{fig:circuit} (a).}
    \label{fig:heavy_hex}
\end{figure}

In this section we introduce dynamic compass codes. We begin by introducing the instantaneous stabiliser group (ISG) formalism, which we will use to explain and study these codes, and then introduce the heavy-hex code and its gauge group. We then show how a carefully chosen measurement schedule for the gauge operators of the heavy-hex code gives rise to the dynamic compass code; a code with constant weight detectors.

\subsection{The ISG formalism}

We make extensive use of the ISG formalism~\cite{Hastings2021, Townsend_Teague_2023} to describe the codes of interest. In this formalism we first define a measurement schedule for a set of Pauli operators and then track the resulting \textit{instantaneous stabiliser group}, i.e. the group of all non-logical Pauli operators that have deterministic measurement outcomes in the case of no noise at a given timestep. When a Pauli operator is measured, the system is now stabilised by that Pauli operator so it is added to the ISG. Any operators already existing in the ISG that anticommute with the measurement are removed. If we measure an operator that already belongs to the ISG then the measurement outcome will be deterministic, and regions of spacetime bounded by measurements whose product is deterministic are called \textit{detectors}. The outcome of the detector can differ from the expected outcome only if an error occurred within it, giving us a way to pinpoint error events in spacetime. In practice it is desirable for detectors to be formed from only a small number of physical measurements so that errors in these measurements do not overly impact our ability to infer the locations of errors affecting the physical qubits

\begin{figure}
    \centering
    \begin{tikzpicture}
        \node at (0,.6) {\includegraphics[width=.2\linewidth]{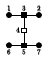}};
        \node at (4.2,0) {
            \begin{adjustbox}{width=.75\linewidth}
            \begin{quantikz}[row sep=.1cm,column sep=.1cm]
                \lstick{1} & \targ{} & & \targ{} & & & & & & & \ctrl{2} & \\
                \lstick{2} & & & & \targ{} & & \targ & & & & \ctrl{1} & & \\
                \lstick{$\ket{0/1}~3$} & \ctrl{-2} & \targ{} & \ctrl{-2} & \ctrl{-1} & \targ{} & \ctrl{-1} & & & \targ{} & \targ{} & \meter{} \\
                \lstick{$\ket{0/1}~4$} & \gate{H} & \ctrl{-1} & \ctrl{1} & & \ctrl{-1} & & \ctrl{1} & & \gate{H} & & \meter{} \\
                \lstick{$\ket{0/1}~5$} & & \ctrl{1} & \targ{} & \ctrl{1} & & \ctrl{2} & \targ{} & \ctrl{2} & \targ{} & \targ{} & \meter{} \\
                \lstick{6} & & \targ{} & & \targ{} & & & & & \ctrl{-1} & & \\
                \lstick{7} & & & & & & \targ{} & & \targ{} & & \ctrl{-2} & \\
            \end{quantikz}
            \end{adjustbox}
        };
        \node at (-1,1.6) {(a)};
        \node at (1.4,1.6) {(b)};
    \end{tikzpicture}
    \caption{(a) The region of the lattice in the yellow box in Fig.~\ref{fig:circuit} (b). (b) A measurement circuit for a plaquette of the heavy-hex lattice as proposed in~\cite{harper2025characterising}. Operationally the circuit measures an $X$ check and its two adjacent $Z$ checks. The circuit is designed such that all measurements are conducted simultaneously to reduce idling time over sequential rounds of measurements. The circuit functions identically regardless of whether qubits $3$, $4$ and $5$ are prepared in the state $\ket{0}$ or $\ket{1}$, removing the need for post-measurement resets.}
    \label{fig:circuit}
\end{figure}
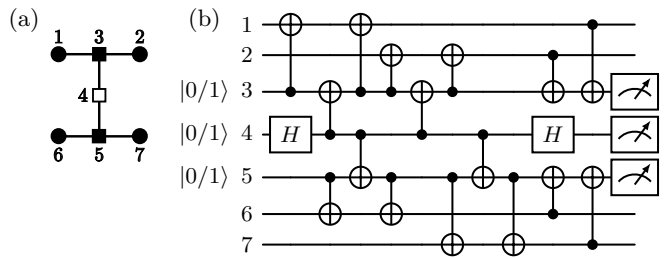

\subsection{The heavy-hex code and its gauge group}

The codes we study in this work are obtained from measurements of operators in the gauge group 
originally defined for the heavy-hex code~\cite{chamberland2020}; an example of a compass code model~\cite{Li2019}. The gauge group of the heavy-hex code is generated by weight-$4$ $X$ operators and weight-$2$ $Z$ operators associated to the faces and edges of a square lattice as shown in~\cref{fig:heavy_hex} (a). These operators can be naturally measured on a device with heavy-hex lattice connectivity~\cite{Chamberland2020topological, sundaresan2022, harper2025characterising, ibm2021heavyhexlattice}. In \cref{fig:circuit} we show a circuit that simultaneously measures a plaquette and its two adjacent edge terms using the qubit connectivity of the heavy-hex lattice architecture~\cite{harper2025characterising}.

\begin{figure}
    \centering
    \includegraphics[width=\linewidth]{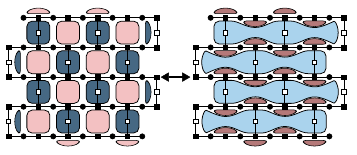}
    \caption{Alternating $X \leftrightarrow Z$ measurement schedule defining the heavy-hex code. Dark blue/red regions show $X$/$Z$ operators measured at that step. Light blue/red regions show $X$/$Z$ operators remaining in the ISG from previous steps.}
    \label{fig:heavy_hex2}
\end{figure}

The heavy-hex code~\cite{chamberland2020} is obtained with a standard measurement schedule where we first measure all $X$ checks, then all $Z$ checks, and repeat. This schedule and the associated transformation of the ISG is shown in~\cref{fig:heavy_hex2}

\subsection{Dynamic compass code}
\label{subsection:dynamic-compass-code}

\begin{figure}
    \centering
    \includegraphics[width=\linewidth]{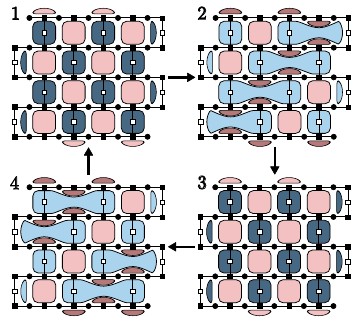}
    \caption{A period-4 schedule that leads to reduced stabiliser weights relative to the heavy-hex code. In steps $1$ and $3$ all $X$ checks are measured and in steps $2$ and $4$ subsets of $Z$ checks are measured. Colour conventions are as in \cref{fig:heavy_hex2}.}
    \label{fig:example_schedule}
\end{figure}

\begin{figure}
\includegraphics[width=.8\linewidth]{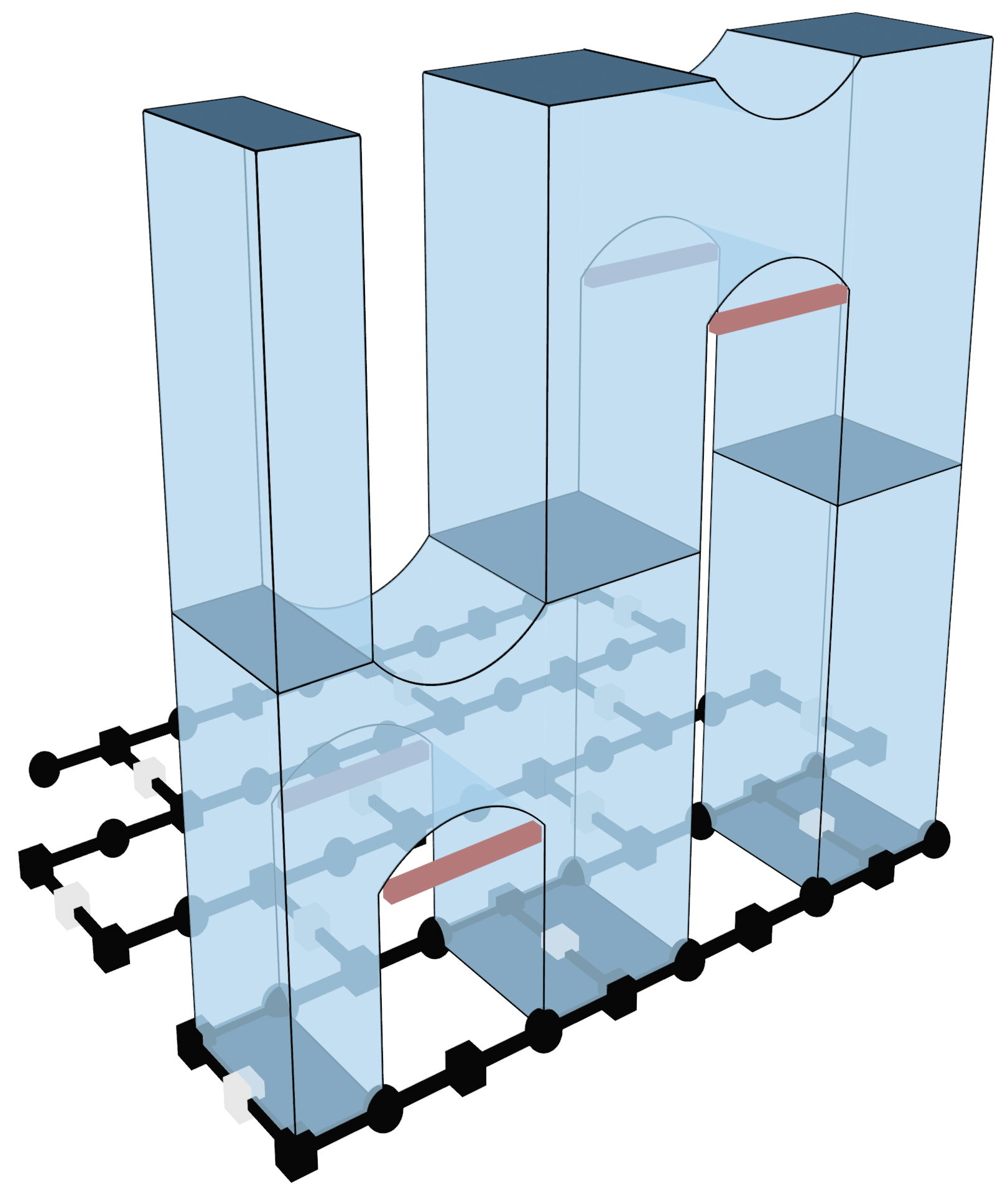}
\caption{Formation of $X$ detectors in the bottom row of \cref{fig:example_schedule} over four steps, beginning at step 1. Time advances along the vertical axis. Dark blue squares show four-qubit $X$ measurements and dark red lines show two-qubit $Z$ measurements. Light blue surfaces show the transformation of $X$ operators in the ISG over time. Spacetime volumes bounded by $X$ measurements are detectors.}
\label{fig:x_detectors}
\end{figure}

\begin{figure}
\includegraphics[width=.8\linewidth]{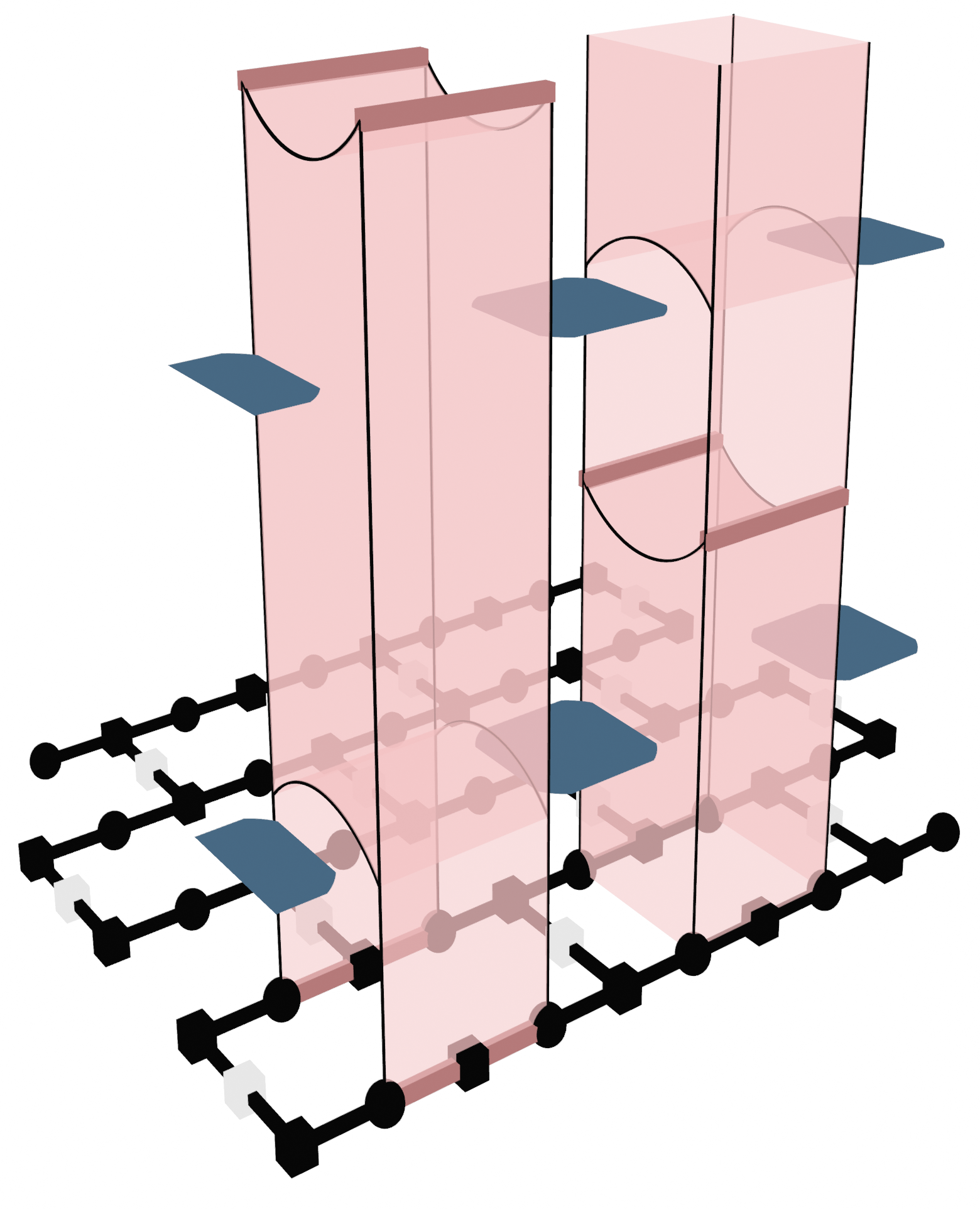}
\caption{Formation of $Z$ detectors in the bottom row and boundary of \cref{fig:example_schedule} over four steps, beginning at step 2. Conventions for visualization are as in \cref{fig:x_detectors}.}
\label{fig:z_detectors}
\end{figure}

 The dynamic compass code is obtained by measuring elements of the heavy-hex code gauge group with a novel measurement schedule,
see, e.g.,~\cref{fig:example_schedule}. This schedule has length four, with all $X$ checks of the code being measured on odd-numbered steps and subsets of $Z$ checks being measured on even-numbered steps. Detecting regions generated by this schedule are shown in \cref{fig:x_detectors} and \cref{fig:z_detectors}. Because all $X$ checks are measured twice per cycle the $X$ detecting regions each span half the cycle. Each detector is formed from either two or four measurements; the former corresponding to the case where a previously measured operator is still in the ISG and the latter to the case where a product of two operators is in the ISG but these individual operators are not. On the other hand, $Z$ checks in the bulk are each measured only once per cycle and so the $Z$ detecting regions span the full cycle and each is formed from exactly four measurements. The $Z$ checks on the boundary are measured twice per cycle, so these detecting regions span half the cycle and are formed from only two measurements. Macroscopically the spacetime structure of these detectors is the same as the surface code, with physical $X$ or $Z$ errors having pointlike syndromes separated in space and measurement errors having pointlike syndromes separated in time. Further illustrations of these detectors and the syndromes produced by different types of error are presented in \cref{section:decoding} 

A notable feature of the schedule of the dynamic compass code is that it generates detectors with constant spacetime volume. This is in contrast to the measurement schedule for the heavy-hex code, in which the $Z$ measurements remove all constant-weight $X$ operators from the ISG, leaving only products of $X$ operators across entire rows of the lattice. Each $X$ detector produced by such a schedule is then formed from $O(d)$ measurements, where $d$ is the distance of the code. Therefore, the outcomes of these detectors become increasingly unreliable as the code grows, preventing scalable fault-tolerance. In subsequent sections we numerically investigate the performance of various other modified schedules and observe that this issue of scalability can indeed be resolved by these modifications, although less frequent $Z$ check measurements can also lead to reduced performance against $X$ errors.

\subsection{Constructing detectors without resets}
\label{subsection:resets}

The circuit used in our simulations is shown in \cref{fig:circuit} and was first presented in Ref.~\cite{harper2025characterising}.
This circuit does not require intermediate resets and can be operated with or without them, which can be useful for current superconducting devices where reset operations may introduce additional noise \cite{geher2024}. In particular, removing resets was shown to be beneficial in a recent experiment on a superconducting device with a heavy-hex layout~\cite{harper2025characterising} and so we focus on the no-reset case in our simulations.

In the reset case, each ancilla is reinitialised before measurement, so the ancilla always starts in the same state. Let $s_t^{i}$ denote the outcome of measurement $i$ at time $t$. A detector is then constructed as the parity of a set of such outcomes across spacetime,
\begin{equation}
D = \sum_{(i,t) \in \mathcal{D}} s_t^{i} ,
\end{equation}
where $\mathcal{D}$ denotes the set of measurements defining the detector.

In the no-reset case, ancillas are not reinitialised and the initial state depends on the previous measurement outcome. Let $m_t^{i}$ denote the measurement outcome in this case. Then
\begin{equation}
m_t^{i} = s_t^{i} + m_{t-1}^{i} .
\end{equation}
The value of $s_t^{i}$ can therefore be recovered as
\begin{equation}
s_t^{i} = m_t^{i} + m_{t-1}^{i} .
\end{equation}
Substituting into the detector definition gives
\begin{equation}
D = \sum_{(i,t) \in \mathcal{D}} \left(m_t^{i} + m_{t-1}^{i}\right).
\end{equation}
If the same ancilla is always used to measure a stabiliser, and stabilisers are measured every round, the detector $D$ in the no-reset case compares measurements separated by two time steps,
\begin{equation}
D = s_t + s_{t-1} = m_t + m_{t-2}. 
\end{equation}

In the dynamic compass code, not all checks are measured once every cycle. 
In our example measurement schedule~\cref{fig:example_schedule}, $Z$ checks (except for boundary checks) are measured once per cycle while $X$ checks are measured twice per cycle. So when building $X$ detectors we must compare measurements which are four steps apart, while for most $Z$ detectors we must compare measurements which are eight steps apart.

\section{Decoding}
\label{section:decoding}

Now that we have introduced an example schedule for a dynamic compass code we can examine the detector-error model and spacetime decoding problem induced by this schedule. For illustrative purposes we focus on the schedule from the previous section in the case where the measurement circuit is operated with resets, but also discuss how the intuition presented here generalises to other cases. 

Importantly, as we will see throughout this section, we find that all types of errors in the circuit noise model will violate adjacent detectors in pairs in a spacetime diagram. We can therefore view errors as short string segment with `pointlike' detection events occurring at the end points of these strings.

\subsection{Measurement errors}

\begin{figure}
    \centering
    \includegraphics[width=.85\linewidth]{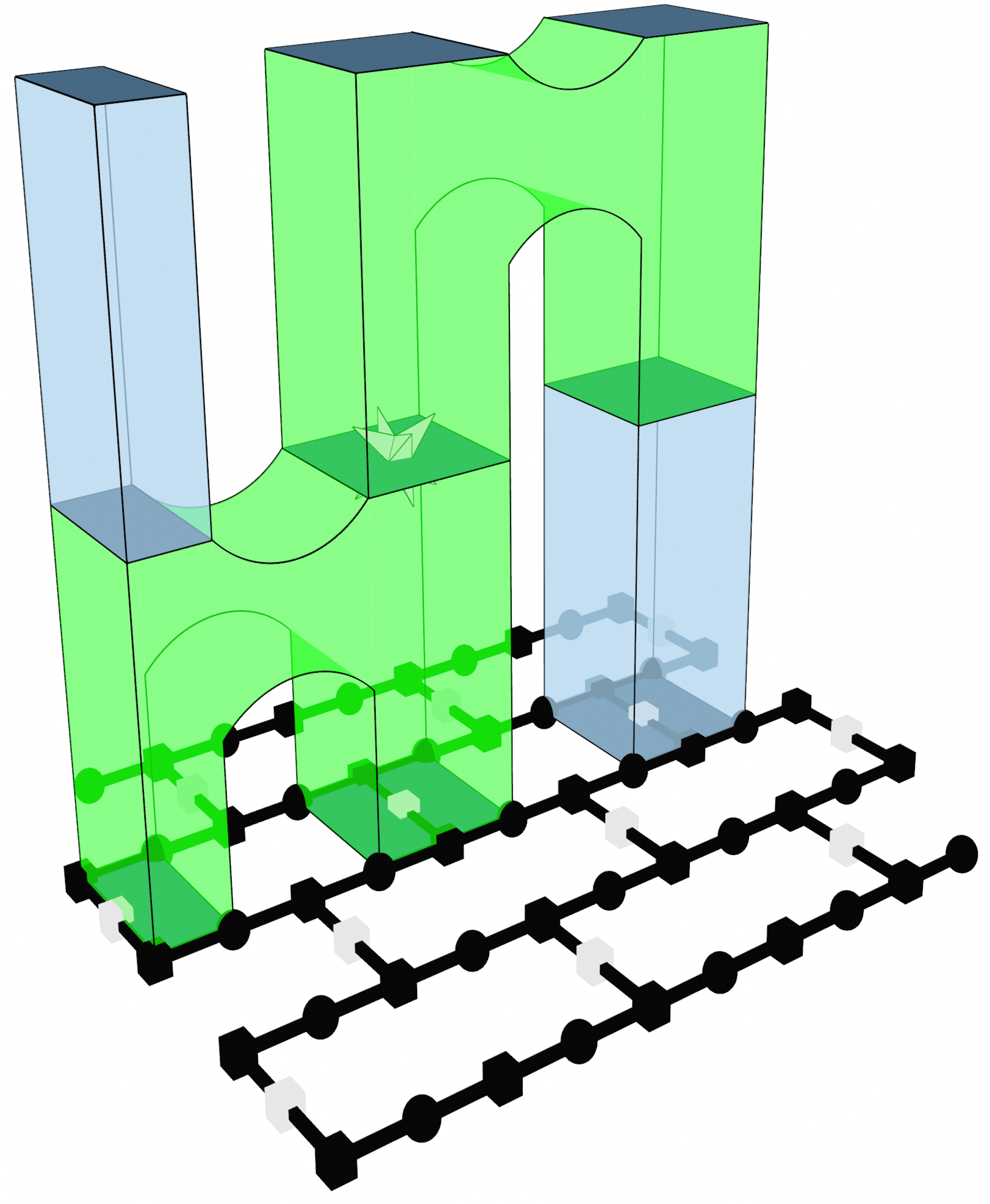}
    \caption{Syndrome of a measurement error (white star) flipping the outcome of a $4$-qubit $X$ measurement. The outcomes of the two green detectors are flipped by the error.}
    \label{fig:x_detector_measurement_error}
\end{figure}

\begin{figure}
    \centering
    \includegraphics[width=.85\linewidth]{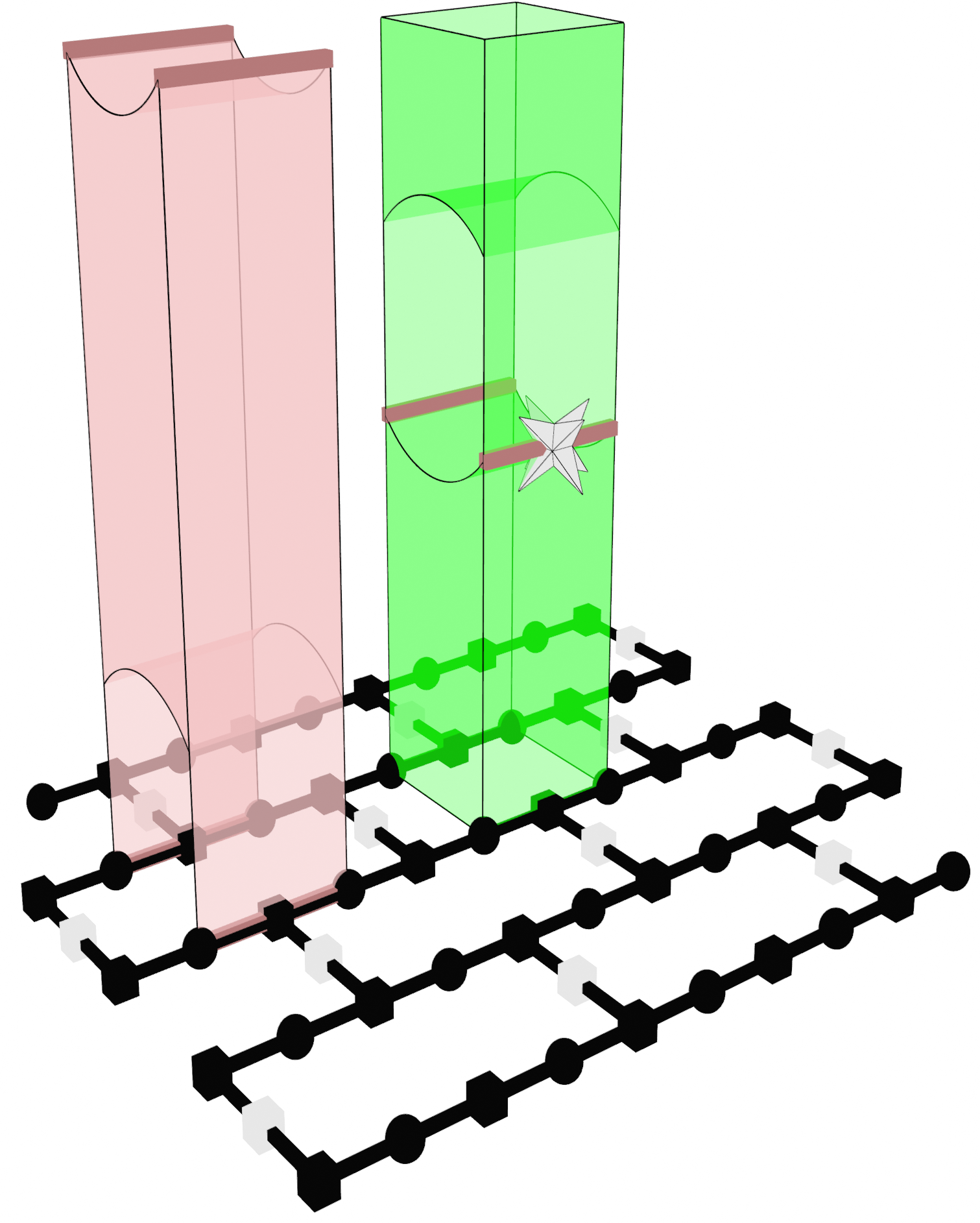}
    \caption{Syndrome of a measurement error (white star) flipping the outcome of a $2$-qubit $Z$ measurement. The outcomes of the two green detectors are flipped by the error.}
    \label{fig:z_detector_measurement_error}
\end{figure}

Because each measurement in this schedule (except those on time-like boundaries) is part of exactly two detectors, errors on these measurements result in pairs of flipped detector outcomes. Examples are shown in \cref{fig:x_detector_measurement_error} and \cref{fig:z_detector_measurement_error}.

All of the schedules we consider in this work consist, like \cref{fig:example_schedule}, of measurements of all $X$ checks on odd steps and subsets of $Z$ checks on even steps. $Z$ checks in the bulk of the code come in pairs such that the product of checks in any pair belongs to the centre of the gauge group, and on any even step of the schedule we always measure either both or neither of the checks in a pair. $Z$ checks on the boundaries themselves belong to the centre of the gauge group and are measured on all even steps. All such schedules also have the property that measurement errors have pointlike syndromes as every measured operator $O_i$ is always either in the ISG or part of a product $\prod_i O_i = O$ where all $O_i$ are measured simultaneously and $O$ is in the ISG. Thus each measured operator is part of a set of measurements that completes an existing detector and begins a new detector, and is not part of any other detector. Notice also that (from the discussion in \cref{subsection:resets}) measurements in the no-reset case also always belong to exacty two detectors, so measurement errors still give rise to pointlike syndromes by the same argument.

\subsection{Data qubit errors}

\begin{figure}
    \centering
    \includegraphics[width=.85\linewidth]{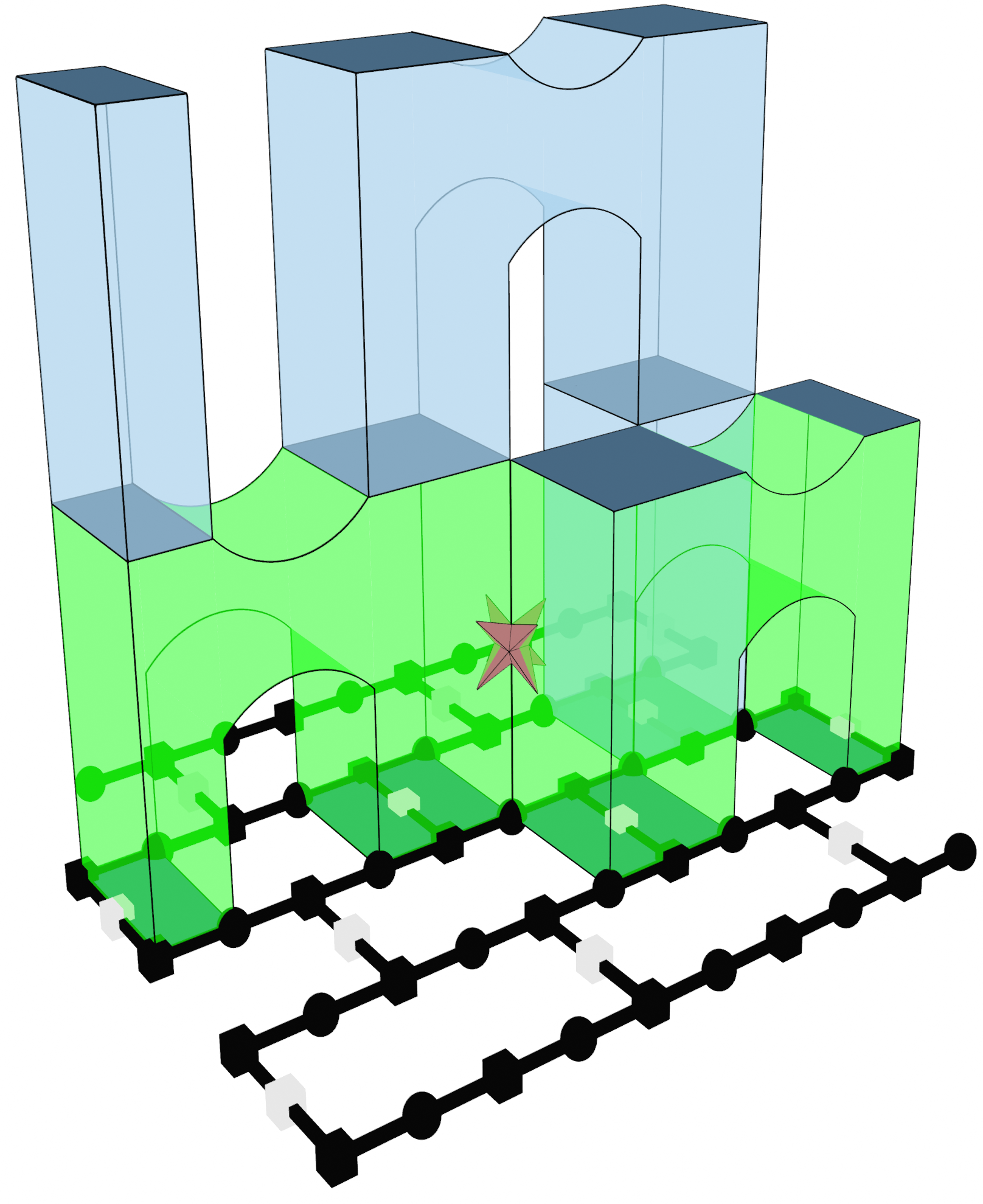}
    \caption{Syndrome of a $Z$ error (red star) occuring on a data qubit. The measurement outcomes of the two green detectors are flipped by the error.}
    \label{fig:detector_z_error}
\end{figure}

\begin{figure}
    \centering
    \includegraphics[width=.85\linewidth]{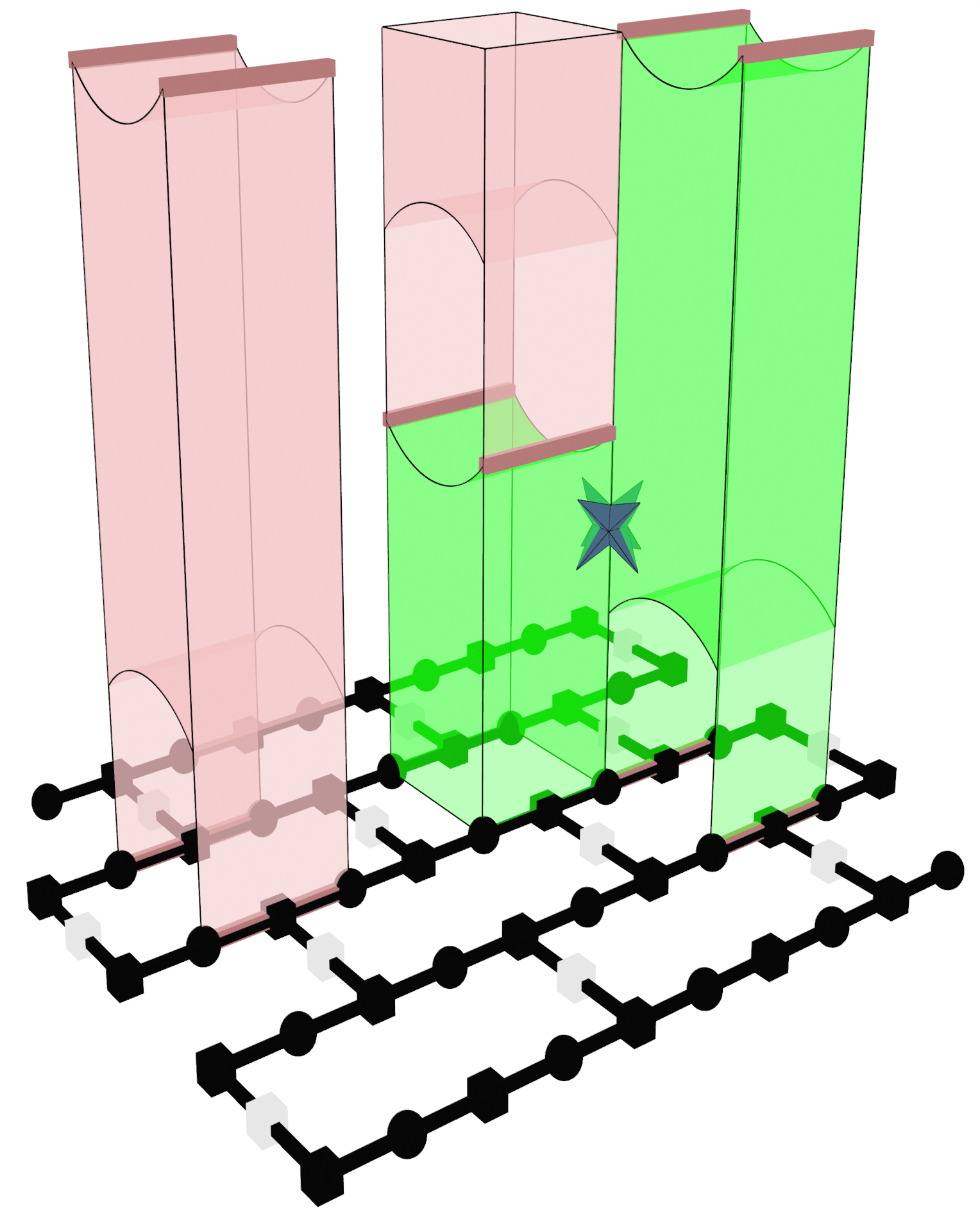}
    \caption{Syndrome of an $X$ error (blue star) occuring on a data qubit. The measurement outcomes of the two green detectors are flipped by the error.}
    \label{fig:detector_x_error}
\end{figure}

As a single detector in this code is formed entirely from either $X$ or $Z$ measurements it detects only $Z$ or $X$ errors (respectively) on data qubits, with $Y$ errors interpreted as the simultaneous occurrence of an $X$ and $Z$ error. As discussed above, each measured operator $O_i$ belongs to two detectors which are disjoint in time. Writing the measurement of $O_i$ at time $t$ as $O_i^t$, an error which anticommutes with this operator occurring between $O_i^{t-1}$ and $O_i^t$ flips the outcome of the detector containing these two measurements, but not the detector containing $O_i^t$ and $O_i^{t+1}$ (or any future detector). Any data qubit in the bulk of the code is in the support of exactly two $X$ or $Z$ generators of the gauge group while qubits on boundaries can be in the support of only a single $X$ or $Z$ generator. We can therefore see that, for any choice of schedule, $X$ or $Z$ errors in the bulk of the code flip detectors in pairs while errors on the boundary flip single detectors. Examples are shown in \cref{fig:detector_z_error} and \cref{fig:detector_x_error}. In contrast to the measurement error case, the flipped $X$ detectors are related by a pure space translation while the flipped $Z$ detectors are related by a diagonal spacetime translation. Finally, note that the set of measurements flipped by a physical error is independent of whether or not we use resets in the measurement circuit, so everything discussed here is also valid for the no-reset case. 

\section{Results}
\label{section:results}

In this section, we present the results of our simulations of our new dynamic compass codes in the no-reset case (for memory). To begin, we describe three different schedules, one equivalent to the standard heavy-hex code and two which produce codes with constant-size detectors but a tradeoff in their ability to protect against $X$ and $Z$ errors. While it is known that the heavy-hex code lacks a threshold against $Z$ errors, we observe that both other schedules demonstrate thresholds in both the $X$ and $Z$ bases. We additionally investigate the performance of these schedules with increasing code distance at a fixed physical error rate and observe exponential suppression of the logical error rate. Finally, we perform stability simulations to estimate the likelihood of logical failures during lattice surgery operations with these codes.

All results in the main text use a uniform circuit-level noise model which is described in more detail in \cref{appdx:noise}.
In \cref{app:3d_plot} we additionally investigate the performance of one of these schedules under a range of asymmetric circuit-level noise models and map out correctable and uncorrectable regions in both the reset and no-reset case (as described in \cref{subsection:resets}). All results presented in this work were obtained using \texttt{stim}~\cite{gidney2021stim} to generate and simulate the circuits, and PyMatching~\cite{higgott2021pymatching} as our minimum-weight perfect matching decoder. Each memory simulation of a given circuit runs for $4d$ full cycles of the measurement schedule, and repeats until we have either observed $1000$ logical failures or performed $10^7$ shots. 
\subsection{Schedules}

\begin{figure*}[t]
    \begin{tikzpicture}
        \node at (0,3) {\Large Schedule};
        \node at (7,3) {\Large $X$ basis};
        \node at (11.5,3) {\Large $Z$ basis};

        \node at (-4,0) {\Large A};
        \node at (0,0) {\includegraphics[width=.4\textwidth]{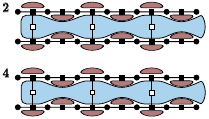}};
        \node at (9,0) {\includegraphics[width=.6\textwidth]{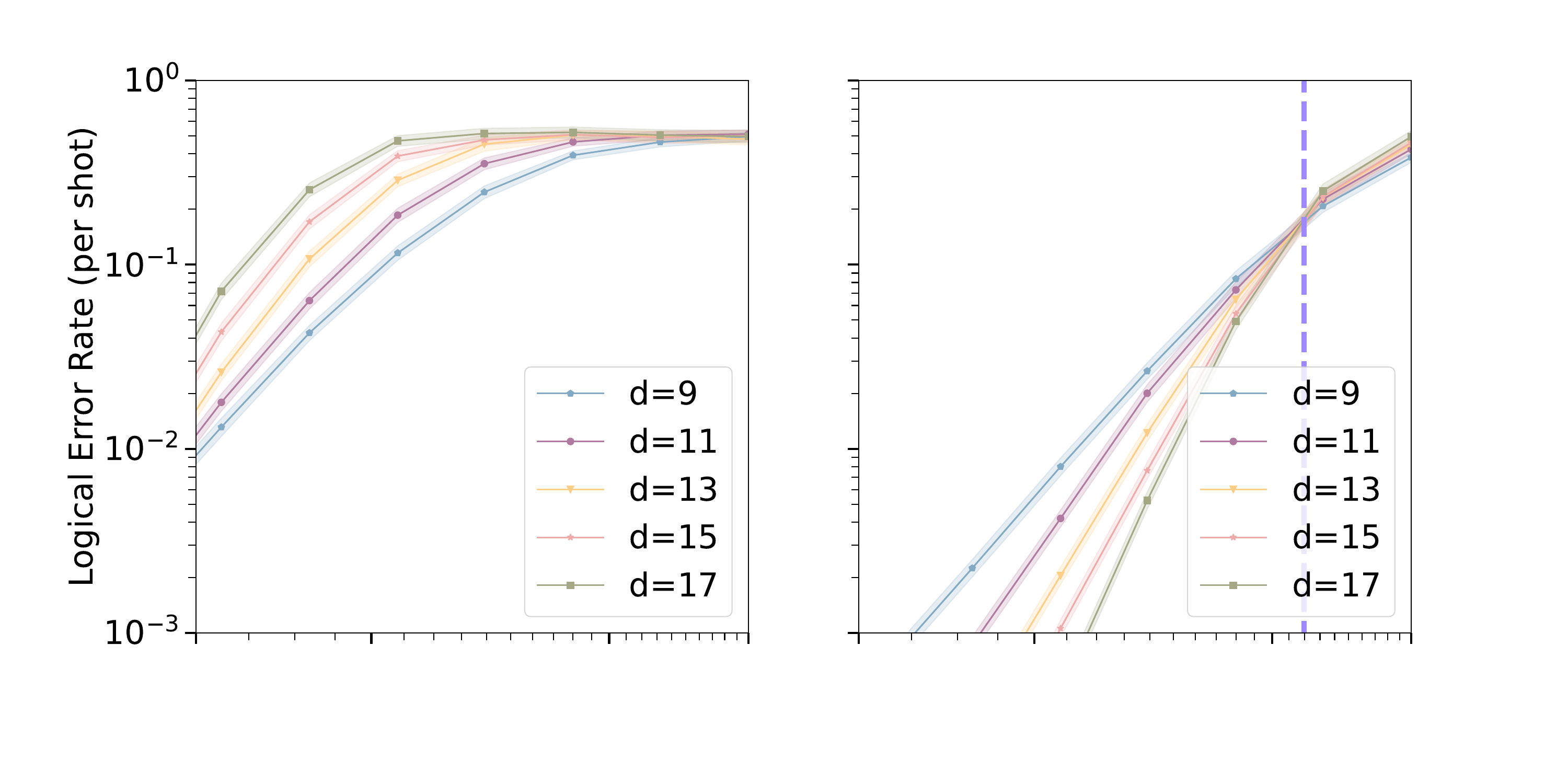}};

        \node at (-4,-4.5) {\Large B};
        \node at (0,-4.5) {\includegraphics[width=.4\textwidth]{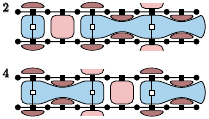}};
        \node at (9,-4.5) {\includegraphics[width=.6\textwidth]{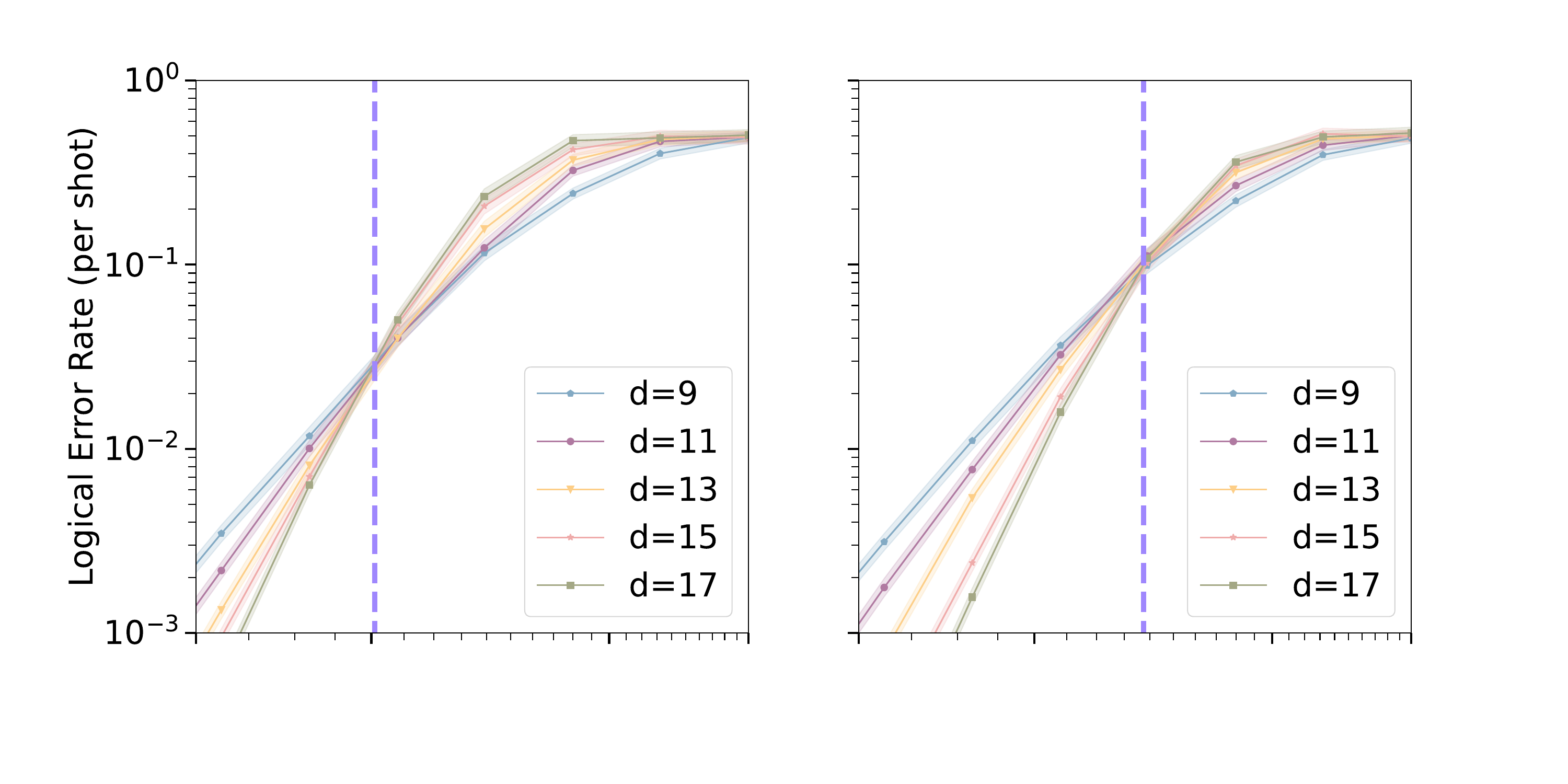}};

        \node at (-4,-9) {\Large C};
        \node at (0,-9) {\includegraphics[width=.4\textwidth]{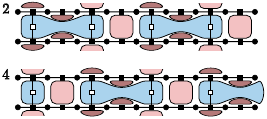}};
        \node at (9,-9) {\includegraphics[width=.6\textwidth]{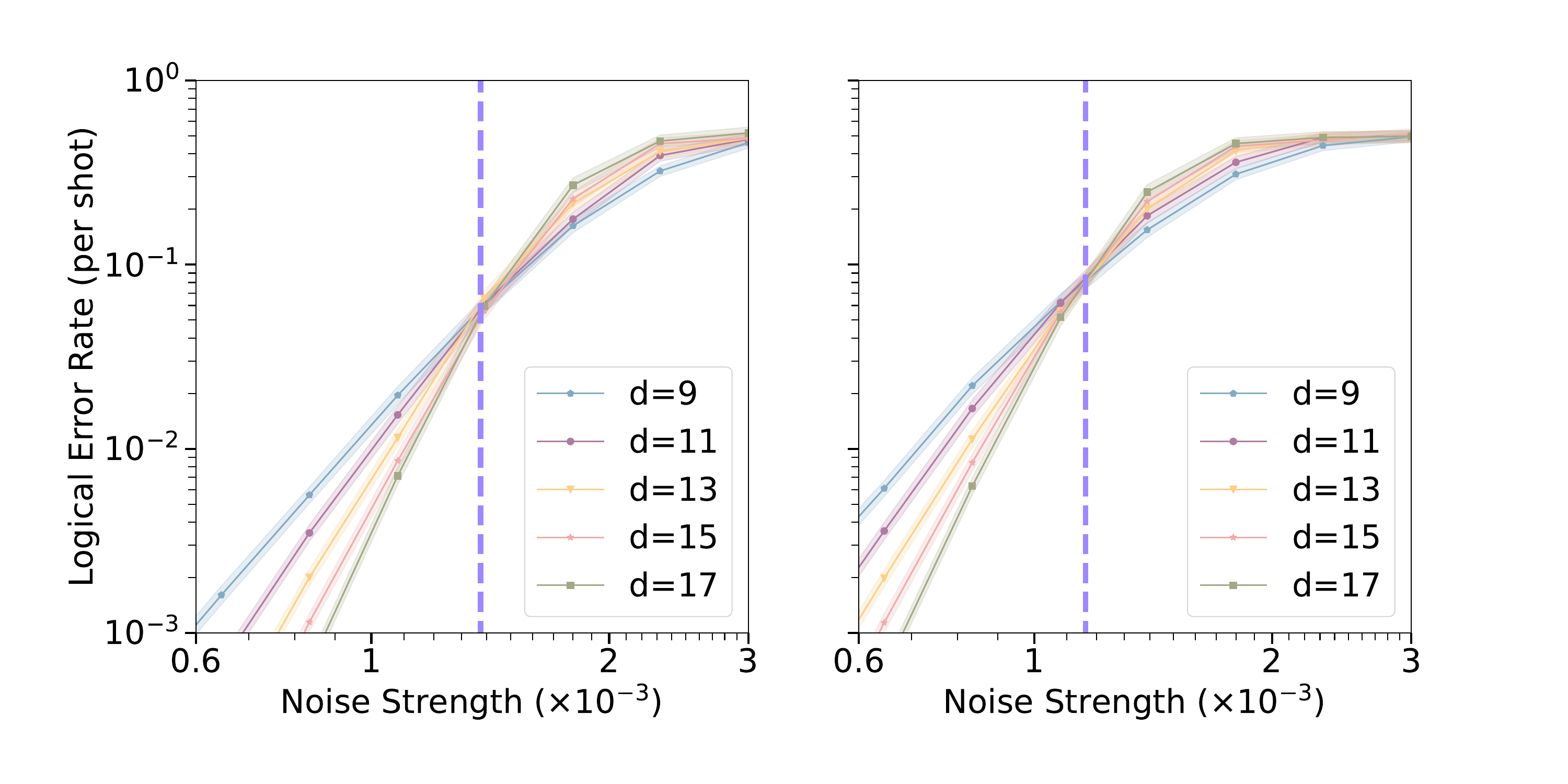}};
    \end{tikzpicture}
    
    \caption{Three different measurement schedules (A, B and C) for one row of $Z$ checks in a dynamic compass code and corresponding performance in the $X$ (middle) and $Z$ (right) bases. Shaded regions represent confidence intervals from \texttt{sinter}~\cite{gidney2021stim} (indicating a Bayes factor of 1000). Each schedule measures different subsets of $Z$ checks on steps $2$ and $4$ (shown). All $X$ checks are measured on steps $1$ and $3$ (not shown) similar to the example schedule in \cref{fig:example_schedule}. $X$ stabilisers in the ISG at these steps are also shown. These row schedules can be tiled and truncated to cover lattices of arbitrary size. Estimated threshold locations are marked in the plots with blue lines. Schedule A corresponds to the standard heavy-hex code and therefore does not demonstrate a threshold in the $X$ basis. Schedule B results in larger $X$ detectors than schedule C but measures some $Z$ checks more frequently, resulting in a tradeoff in performance against $X$ vs $Z$ errors. Indeed, we observe that the threshold in the $X$-basis increases as we remove more $Z$-measurements at each step. This is expected because we have more X-basis stabilisers, and they have lower weight. In contrast, we see that the threshold for errors detected by $Z$-basis measurements decreases as we remove more $Z$-basis measurements from each schedule. This is expected because $Z$-basis stabilisers are being measured less frequently across schedules A through C.}
    \label{fig:schedules}
\end{figure*}
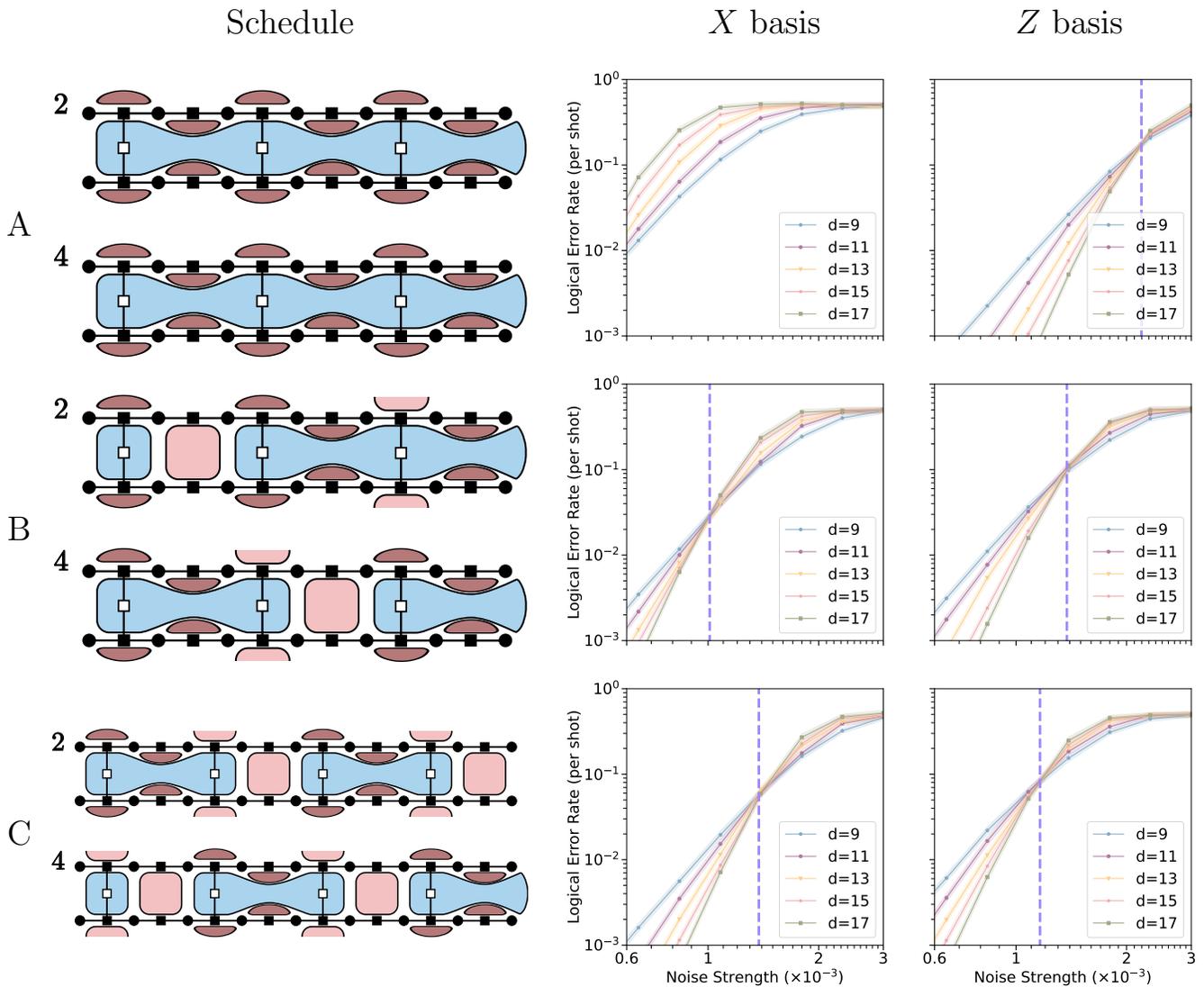

We compare three measurement schedules in the dynamic compass code in \cref{fig:schedules}. The difference between these schedules is the choice of $Z$ measurements that are omitted at different time steps. Let us assume that each full measurement round consists of four steps: $X$ measurements are performed at steps $1$ and $3$ as in \cref{fig:example_schedule}, while different choices of $Z$ measurements are performed at steps $2$ and $4$. Since $X$ measurements are always performed, we only explicitly indicate the timing of the $Z$ measurements.   

Here we show three different schedules: the original heavy-hex code with no omitted $Z$ measurements (schedule A), a schedule omitting one in every three $Z$ measurements along each row per step (schedule B), and a schedule omitting alternate $Z$ measurements along each row on each step (schedule C). We observe that increasing the number of omitted $Z$ measurements improves the threshold in the $X$ basis while degrading the threshold in the $Z$ basis. This trade-off arises because omitting $Z$ measurements reduces the weights of $X$ detectors and improves our ability to correct $Z$ errors. However, fewer $Z$ measurements also reduce the available syndrome information per step, leading to an increase in the spacetime volume of $Z$-detectors and causing worse performance in this basis.

This trade-off reflects a general design principle: reducing stabiliser weight in one basis necessarily increases the effective spacetime extent of detectors in the conjugate basis. This behavior is consistent with the mechanism discussed in \cref{section:prelim}, where omitting $ZZ$ gauge measurements preserves intermediate $Z$ stabilisers and segments long $X$ stabilisers into smaller components. To achieve this effect, the omitted $ZZ$ gauge measurements must be those that would otherwise factorize the intermediate four-qubit $Z$ stabilisers into two-qubit operators. Omitting other $ZZ$ measurements that do not affect these stabilisers has little impact on bounding the $X$ stabiliser weight. At a more global level, maintaining stable $Z$-error suppression requires that $Z$ measurements be omitted in a uniform pattern across each row. If the omission is not evenly distributed, the stabiliser weight can only be bounded locally rather than across the entire lattice. In addition, the omission pattern must be periodic in time, so that intermediate $Z$ stabilisers are only preserved over constant measurement rounds. When these conditions are satisfied, the resulting detector sizes remain bounded in both the $X$ and $Z$ bases, leading to a more balanced and stable overall performance.

\begin{figure}
    \begin{tikzpicture}
        \node at (0,0) {\includegraphics[width=\linewidth]{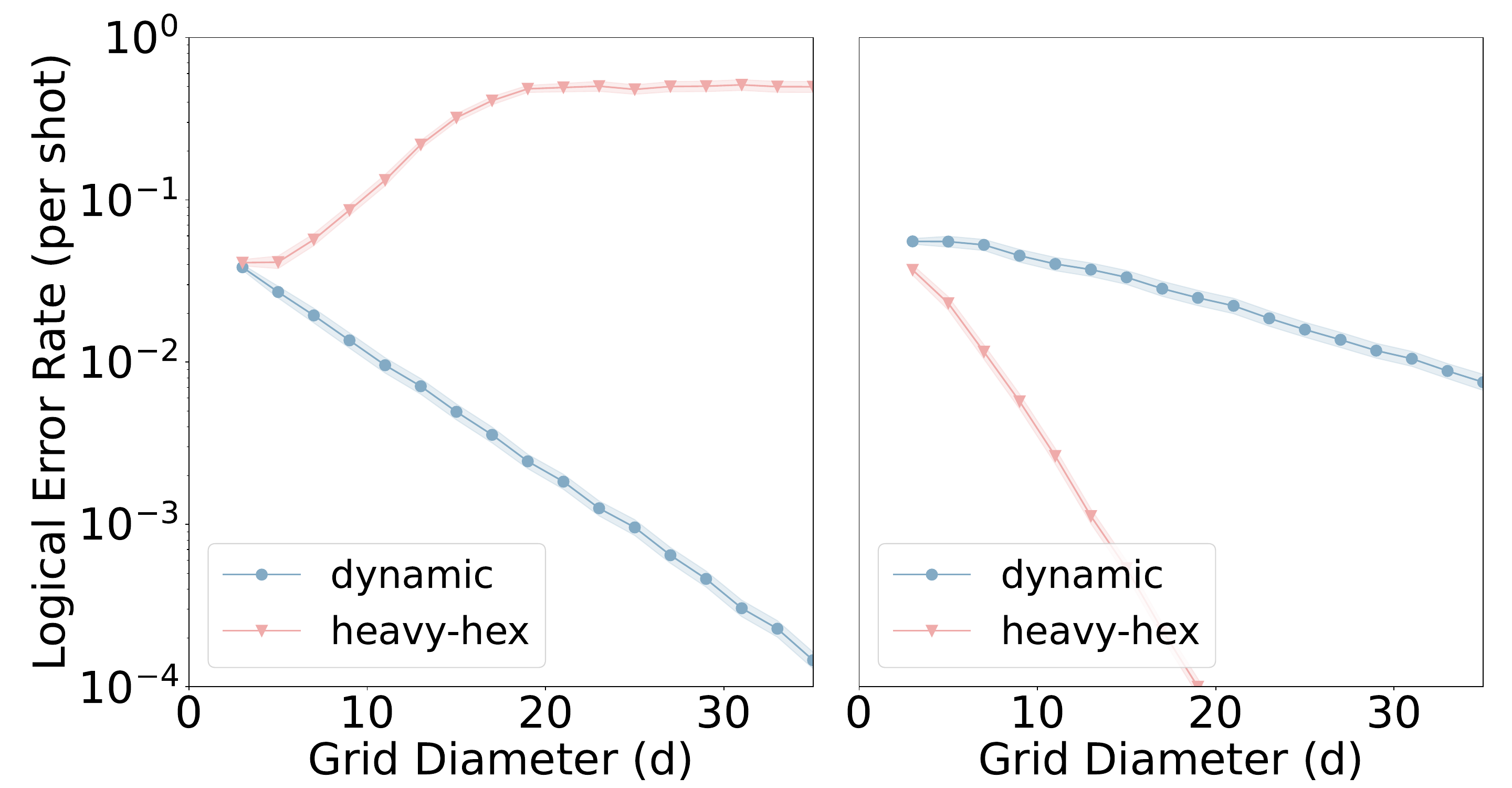}};
        \node at (-4,2.3) {(a)};
        \node at (.4,2.3) {(b)};
        \node at (-1.5,2.4) {\large $X$ basis};
        \node at (2.4,2.4) {\large $Z$ basis};
    \end{tikzpicture}
    
    \caption{(a) $X$ basis and (b) $Z$ basis logical error rates for schedules A and C from \cref{fig:schedules} under uniform circuit-level noise with strength $p=10^{-3}$. Shaded regions represent confidence intervals from \texttt{sinter}~\cite{gidney2021stim}. While increasing the distance of the code monotonically decreases the logical error rate of the heavy-hex code in the $Z$ basis, the logical error rate plateaus in the $X$ basis. The dynamic schedule allows the arbitrary suppression of the logical error rate in both bases.}
    \label{fig:various-d}
    
\end{figure}
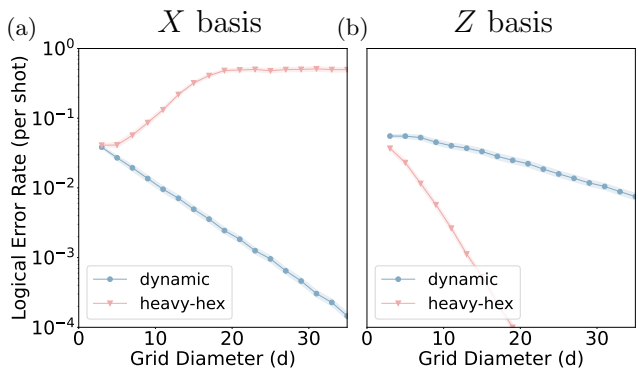

The original heavy-hex code is known to lack a threshold in the $X$ basis. This is consistent with what we observe in  \cref{fig:schedules} and also in \cref{fig:various-d}, where increasing the code distance does not reduce the logical error rate. In contrast, for schedule C we observe exponential suppression of the logical error rate with increasing code distance in both bases, although performance in the $Z$ basis is reduced relative to the heavy-hex code.

\subsection{Stability Performance}

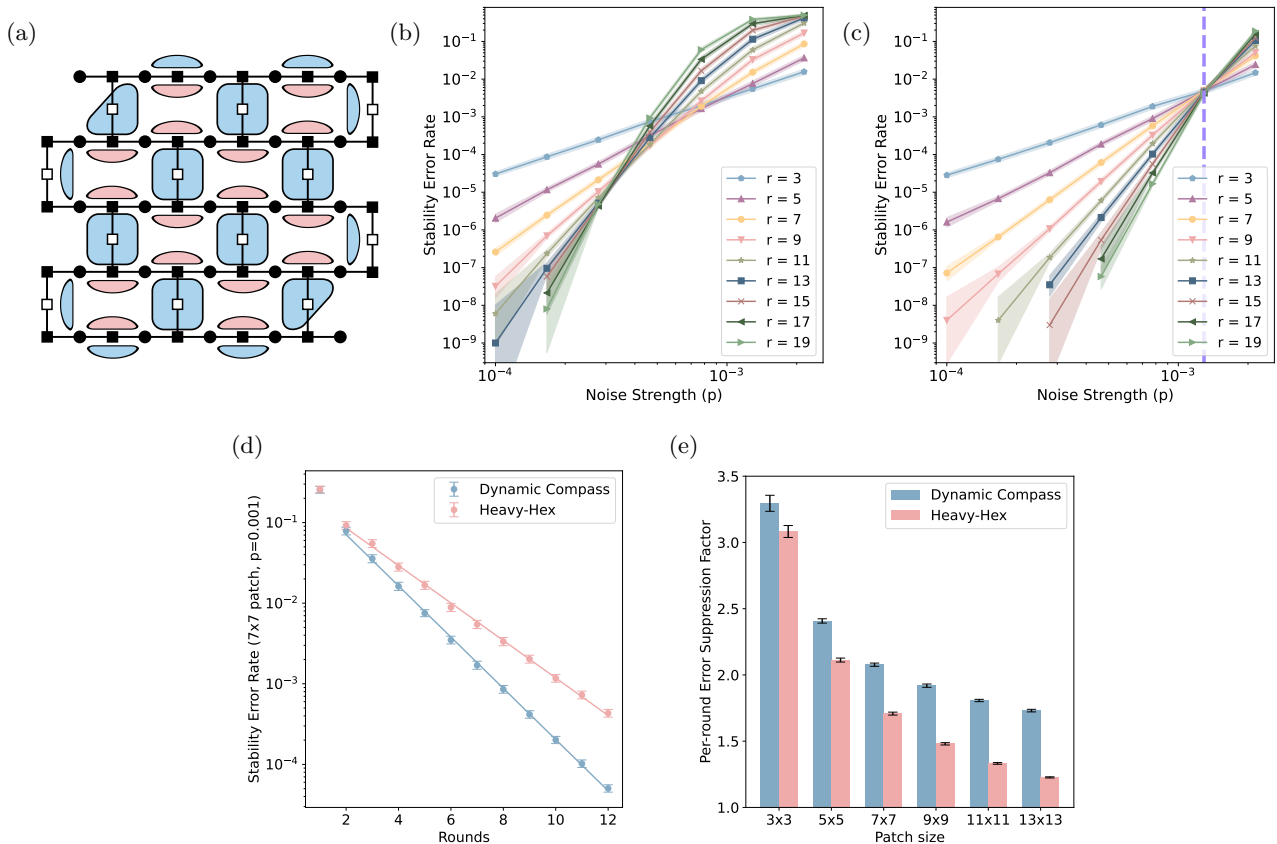
\begin{figure*}[ht]
\begin{tikzpicture}
    \node at (0,0) {\includegraphics[width=0.25\linewidth]{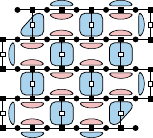}}; 
    \node at (5.5,0) {\includegraphics[width=0.32\linewidth]{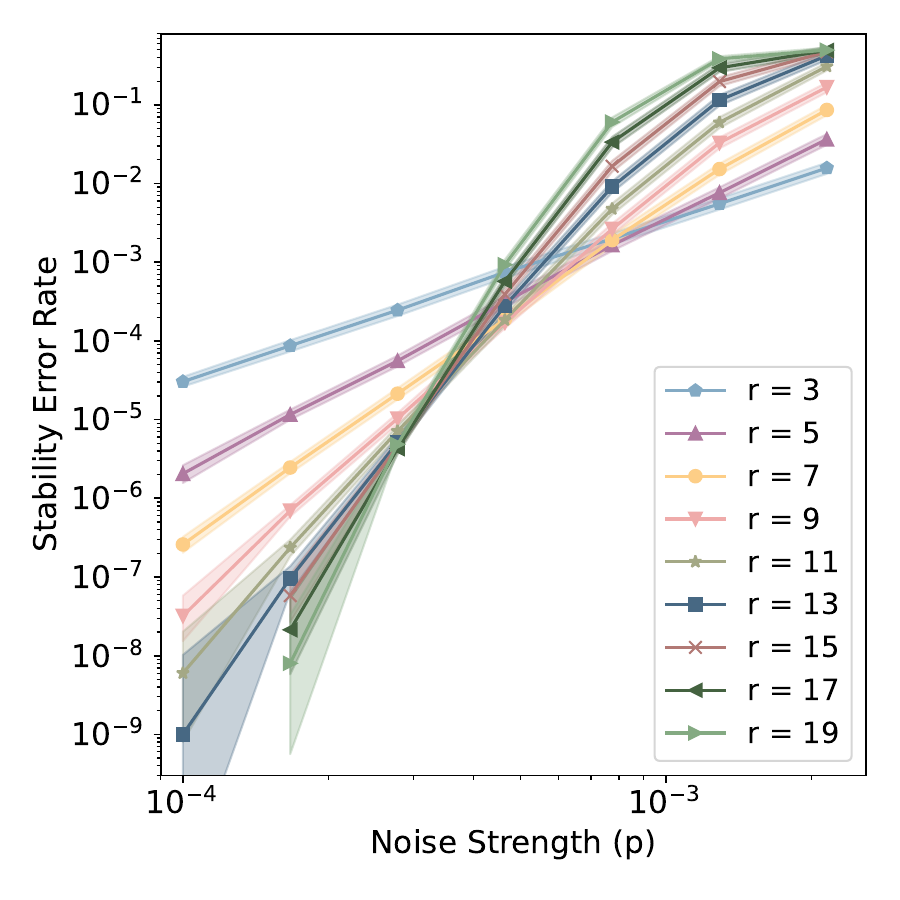}};
    \node at (11.5,0) {\includegraphics[width=0.32\linewidth]{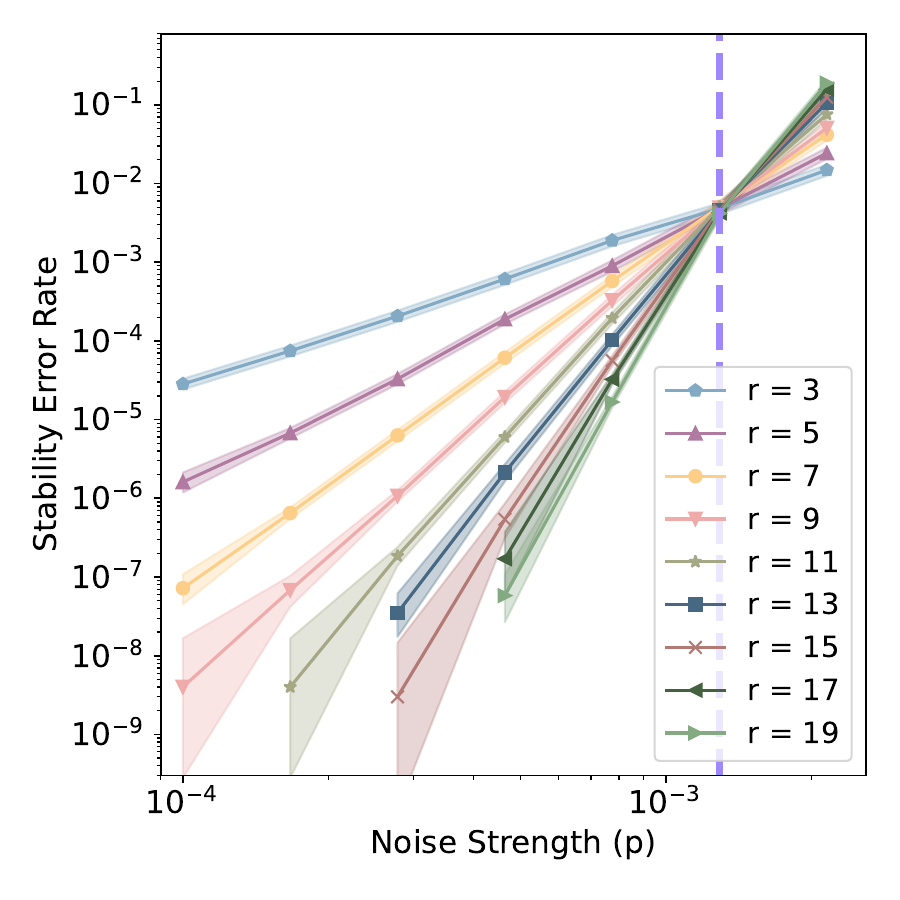}};
    \node at (3,-6) {\includegraphics[width=0.3\linewidth]{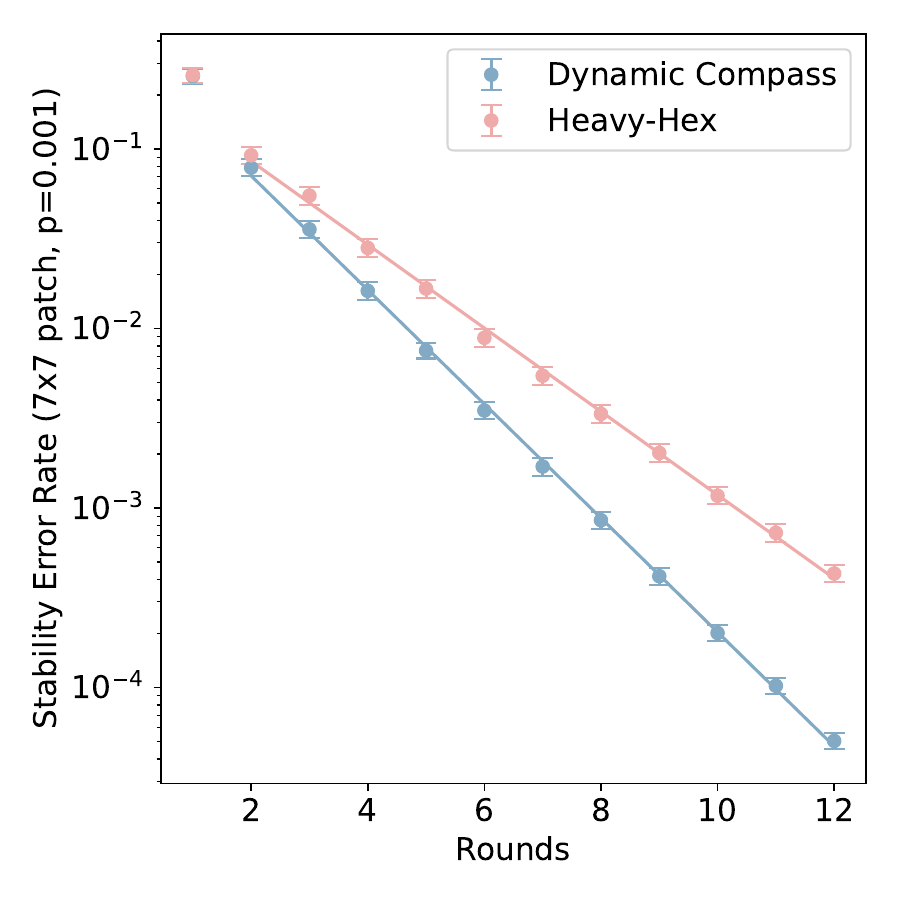}}; 
    \node at (9,-6) {\includegraphics[width=0.3\linewidth]{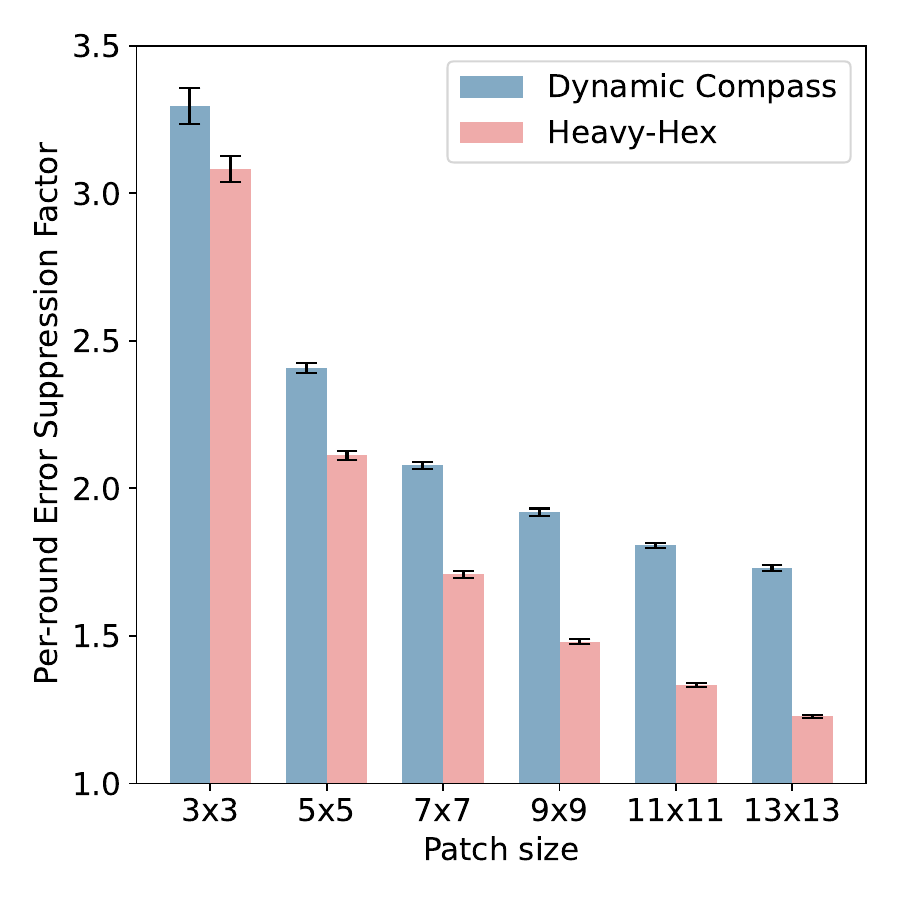}};
\node[inner sep=0pt] at (-2.5,2.3)         {(a)};
\node[inner sep=0pt] at (2.6,2.3)         {(b)};
\node[inner sep=0pt] at (8.6,2.3)         {(c)};
\node[inner sep=0pt] at (0.5,-3.2)         {(d)};
\node[inner sep=0pt] at (6.3,-3.2)         {(e)};
\end{tikzpicture}
    \caption{Comparison of stability performance between the dynamic compass and heavy-hex codes. (a) An example of a $5\times 5$ stability patch used for both codes. Boundary conditions are altered so that the parity of all blue $X$-checks is always even, in the absence of errors. (b) Stability ``threshold plot" for the heavy-hex code shows the absence of a threshold. Each line corresponds to a stability experiment run on an $r\times r$-sized patch for $r$ rounds. (c) Stability threshold plot for the dynamic compass code displays a clear threshold, indicated with a dashed line. (d) The logical error rate for a stability experiment decays exponentially in the number of rounds for both codes, but the error suppression factor is larger for the dynamic compass code. Data is shown for a $7\times 7$ patch with a noise strength of $p=10^{-3}$. (e) Comparison of per-round stability error suppression factors between the two codes, for patches of various sizes, and for a noise strength of $p=10^{-3}$. All uncertainties indicate a Bayes factor of 1000, apart from in (e), where error bars indicate the (propagated) standard deviation in the linear fits.\label{fig:stability_results}}
\end{figure*}

The heavy-hex code also lacks a threshold for stability experiments, which additionally limits its scalability when performing measurement-based logic gates. Stability experiments are the `dual' of memory experiments~\cite{Gidney2022dual, harper2025characterising}. They measure the ability for a quantum error-correcting code to fix a Pauli frame, which is crucial for various tasks in fault-tolerant computation, including changes to a code patch or lattice surgery. To set up this experiment, we initialize a $d\times d$-sized patch with modified boundary checks, an example of which (for $d=5$) is shown in \cref{fig:stability_results}(a). The product of all blue $X$-checks in this patch is the identity, and so the product of all these measurements constitutes a logical observable that should be $+1$ in the absence of noise. The initial data qubit resets are performed in the $Z$ basis (and, similarly, the final readout measurements are in the $Z$ basis) so that each $X$-check individually has a random outcome, while the product of all such checks is deterministic. We include mid-circuit resets in these simulations to maintain maximal time-like distance~\cite{geher2024}.

The heavy-hex code does not demonstrate a threshold for this stability experiment: for any fixed noise pararameter $p$, as we grow all dimensions of the experiment's spacetime volume (the size of the patch, $d$, and the number of syndrome extraction rounds, $r$, which we set equal to $d$), the error rate in the stability observable eventually increases with $d$. Once again, this is due to the fact that $X$-basis detectors scale with the size of the system. The absence of threshold can be seen from \cref{fig:stability_results}(b), where the crossing points between successive curves shift to the left. This results from the large-weight $X$-detectors, which contain an extensive number of measurements, and so suffer from an error detection rate approaching $50\%$ as the size of the patch is increased.

The dynamic compass code has a stability threshold because the high-weight $X$-detectors are broken up into detectors of constant volume. The threshold can be seen in \cref{fig:stability_results}(c) at $\approx 1.3\times 10^{-3}$, where all curves intersect. We test stability experiments with round numbers from $r=3$ to $r=19$ and construct a schedule similar to that shown in \cref{fig:example_schedule}, with $Z$ detectors along a row alternating between being measured and being skipped. 

As we increase the number of rounds for a fixed-size patch, the logical error rate decreases. In addition to evaluating a threshold, it is also important to determine the degree of logical error rate suppression per round for the stability experiment. That is, the factor $s(d,p)$ (for a $d\times d$ patch with noise strength $p$) by which the error rate of the logical observable of the stability experiment decreases as we increase the number of rounds $r$ by $1$: $\epsilon(r+1)\approx \epsilon(r)/s(d,p)$. This is (the exponential of) the gradient of the lines plotted in \cref{fig:stability_results}(d), which show an example of the exponential suppression in stability error rate as we increase the number of rounds while keeping the patch size fixed (for $d=7$, $p=10^{-3}$). We see that the per-round suppression factor $s$ is larger for the dynamic code than the heavy-hex code. Observing this difference in suppression factors over a range of patch sizes in \cref{fig:stability_results}, we see that the two codes start off similarly at $d=3$ (where the measurement schedules differ minimally), but that the heavy-hex code approaches a suppression factor of $\approx 1$ (no suppression), while the suppression factor for the dynamic compass code performs comparatively better as we increase the patch size. We expect this suppression factor to approach a constant for the dynamic code, since for sufficiently low $p$, we expect the stability error rate to be $P_S \sim (p/p_\text{th})^r \cdot \text{poly}(d)$, where $p_\text{th}$ is the threshold for stability, and where the polynomial factor accounts for the number of minimum-weight logical errors that exist. In Appendix~\ref{app:stab_error_supp}, we provide evidence for this.

\section{Lattice surgery with dynamic compass codes}
\label{section:surgery}

In addition to protecting quantum states, a useful quantum code should also allow for reliable manipulation of stored information. In surface codes a leading technique is lattice surgery~\cite{Horsman2012} and in most cases this can be straightforwardly extended to the dynamic compass code discussed in this work, as we now show. Lattice surgery in this code is in principle scalable, owing to the existence of thresholds for both memory and stability experiments. We use the schedule from \cref{fig:example_schedule} in these examples but note any of the schedules discussed in this work can also be used. 

\begin{figure}
    \centering
    \includegraphics[width=\linewidth]{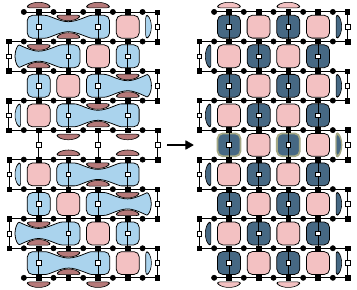}
    \caption{$X$ basis lattice surgery in the dynamic compass code. The product of the $X$ measurements with thick green borders is a logical $\overline{XX}$ of the two code patches.}
    \label{fig:x_surgery}
\end{figure}

The easiest case is parity measurement by lattice surgery in the Pauli-$X$ basis as shown in \cref{fig:x_surgery}. Here we begin with two separate code patches (in step four of \cref{fig:example_schedule}) and join them in the next timestep by measuring new $X$ operators whose product is a logical $\overline{XX}$ operator. These operators then also need to be included in the next $\sim d$ rounds of $X$ measurements in order to infer the logical measurement outcome reliably, which takes $\sim d/2$ cycles of \cref{fig:example_schedule} as there are two $X$ measurement steps per cycle.

\begin{figure}
    \centering
    \includegraphics[width=\linewidth]{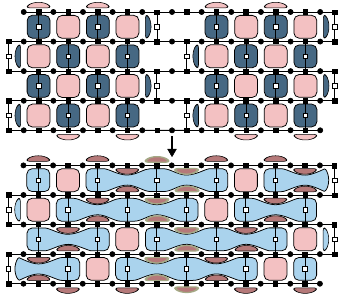}
    \caption{$Z$ basis lattice surgery in the dynamic compass code. The product of the $Z$ measurements with thick green borders is a logical $\overline{ZZ}$ of the two code patches.}
    \label{fig:z_surgery}
\end{figure}

We also consider a logical parity measurement in the Pauli-$Z$ basis by lattice surgery, see~\cref{fig:z_surgery}. We start with two code patches at step one of \cref{fig:example_schedule} and note that these must be slightly further apart than in the previous case so that each boundary stabiliser can be associated with a unique ancilla qubit. In the next timestep we measure new $Z$ operators whose product is a logical $\overline{ZZ}$ operator. This requires measurement of all $ZZ$ pairs in both columns between the patches and so creates larger-weight $X$ stabilisers between the patches than are otherwise admitted by the schedule, but this weight is still independent of $d$ and so the corresponding detectors will still be constant size. These new pair measurements must then be included in the next $\sim d$ $Z$ measurement steps which once again requires $\sim d/2$ measurement cycles. Note that, since $Z$ detectors are formed only once over four measurement rounds in the bulk (one cycle of \cref{fig:example_schedule}), we might expect to halve the time-like distance of a lattice surgery procedure, in comparison to the heavy-hex code. However, the flexibility of the schedule allows us to measure all shaded $Z$ checks in \cref{fig:z_surgery} twice per cycle while maintaining constant-weight detectors, thereby retaining the full time-like distance. 

There also exist techniques for direct measurement of a logical $Y$ operator in the surface code and implementation of $S$ gates~\cite{Gidney_2024_Y_basis}, along with techniques for performing single-qubit Hadamard gates~\cite{Geher_2024_Hadamard}, but adapting these to the heavy-hex lattice is more difficult as they require the creation and movement of twist defects. 
As such, we leave the adaptation of these more complex techniques as a problem for future work.

\section{Discussion}
\label{section:discussion}

We have introduced a family of dynamic compass codes that can be realised with the heavy-hex lattice connectivity using a relatively modest qubit footprint. We evaluate the performance of several instances of readout schedules, and interrogate a tradeoff between the threshold error rate for Pauli-$X$ logical errors and Pauli-$Z$ logical errors with different choices of schedules. We also identify a threshold using simulations of a stability experiment, therefore showing we can scale the code while performing measurement-based logic gates. We evaluate the threshold error rates over a broad range of noise parameters to determine the correctable region in this parameter space. Lastly, we investigate logical entangling operations using this code.

We have obtained our code family by choosing a novel dynamic measurement schedule for the gauge group of the heavy-hex code; an example of a compass code. The use of a dynamic measurement schedule therefore generalises the notion of a compass code where we nominally obtain a subsystem code with a standard measurement schedule. As we have alluded in the introduction of this manuscript, many known examples of dynamic codes could be constructed with a careful choice of measurement schedule for compass codes, including the seminal example of the honeycomb Floquet code introduced by Hastings and Haah~\cite{Hastings2021}. All of these examples then, including the dynamic compass code we have presented here, point to a more general way of constructing dynamic codes using the framework of compass codes. Indeed, we have proposed the term `dynamic compass code' to provoke the reader to explore this method of constructing dynamic codes further.

In an accompanying paper, we provide an experimental demonstration of a distance-5 dynamic compass code on an array of superconducting qubits~\cite{DCCexperiment}.
Looking ahead, to produce a large-scale quantum-computing architecture, we will need to find low-overhead ways of performing logic gates with these codes. Underlying our dynamic syndrome extraction circuit, we have a topological code that is essentially equivalent to a surface code. One way to see this is to check that at any given point in the syndrome extraction circuit, the instantaneous stabiliser group is that of a surface code for some choice of lattice geometry that is varying over time. That means we can expect to be able to perform Clifford gates by manipulating the twist defects that appear at the corners of the code~\cite{Brown2017,Geher_2024_Hadamard} and indeed, as we have shown, we can perform lattice surgery operations. See also Ref.~\cite{ellison2023floquetcodestwist} where it is considered how one can manipulate twists in a dynamic code. It will be interesting if we can perform similar logic gates using our dynamic compass code while respecting the underlying low-valency qubit architecture.

\begin{acknowledgments}
We thank the organisers of the Quantum Error Correction Workshop at the Yukawa Institute for Theoretical Physics where the initial ideas for this project were conceived.
BJB is grateful for the hospitality of the Center for Quantum Devices at the University of Copenhagen.
Numerical results presented in this work were obtained using the HPC resources provided by the Scientific Computing and Data Analysis section of the Research Support Division at OIST. 
This work was supported in part by Japan Science and Technology Agency (JST) as part of Adopting Sustainable Partnerships for Innovative Research Ecosystem (ASPIRE), Grant Number JPMJAP25A3.
J.Z. acknowledges the support of the 2025 Google PhD Fellowship.
We acknowledge support from the Intelligence Advanced Research Projects Activity (IARPA), under the Entangled Logical Qubits program through Cooperative Agreement Number W911NF-23-2-0223 (R.H., B.J.B., and S.D.B.).  The views and conclusions contained in this document are those of the authors and should not be interpreted as representing the official policies, either expressed or implied, of IARPA, the Army Research Office, or the U.S. Government. The U.S. Government is authorized to reproduce and distribute reprints for Government purposes notwithstanding any copyright notation herein. 
S.H.L was supported by the National Research Foundation of Korea (NRF) grant funded by the Korea government (MSIT) (RS-2026-25476454) and supported by Creation of the Quantum Information Science R\&D Ecosystem (Grant No. RS-2023-NR068116) through the National Research Foundation of Korea (NRF) funded by the Korean government (Ministry of Science and ICT).

\end{acknowledgments}


\input{main.bbl}

\appendix
\section{Noise model}
\label{appdx:noise}

Most circuits in this paper were simulated using the uniform noise model summarized in Table~\ref{tab:noise-channels} and Table~\ref{tab:ideal-noisy-gates}. Table~\ref{tab:noise-channels} defines the individual noise channels (\(\text{MERR}_B\), \(\text{XERR}(p)\), \(\text{ZERR}(p)\), \(\text{DEP1}(p)\), \(\text{DEP2}(p)\)) used to build up noisy versions of gates. Table~\ref{tab:ideal-noisy-gates} then shows how each ideal gate is replaced by its noisy counterpart under this model. 

\begin{table}[ht]
    \centering
    \caption{Definitions of noise channels used to define noisy versions of gates in Table~\ref{tab:ideal-noisy-gates}.}
    \label{tab:noise-channels}
    \renewcommand{\arraystretch}{1.2}  
    \begin{tabular}{l p{5cm}}
    \hline
    \textbf{Noise Channel} & \textbf{Probability Distribution of Effects} \\
    \hline
    \(\text{MERR}_B(p)\) & 
    \((1-p) \;\rightarrow\; M_B \) \newline
    \(\quad p \;\rightarrow\; M_{-1 \cdot B}\) (i.e., measurement result is inverted) \\
    \hline
    \(\text{XERR}(p)\) & 
    \((1-p) \;\rightarrow\; I \) \newline
    \(\quad p\;\rightarrow\; X \) \\
    \hline
    \(\text{ZERR}(p)\) & 
    \((1-p) \;\rightarrow\; I \) \newline
    \(\quad p\;\rightarrow\; Z \) \\
    \hline
    \(\text{DEP1}(p)\) & 
    \( (1-p) \;\rightarrow\; I \) \newline
    \(\quad \tfrac{p}{3} \;\rightarrow\; X, \quad \tfrac{p}{3} \;\rightarrow\; Y, \quad \tfrac{p}{3} \;\rightarrow\; Z \) \\
    \hline
    \(\text{DEP2}(p)\) & 
    \( (1-p) \;\rightarrow\; I \otimes I\) \newline
    \(\quad \tfrac{p}{15} \;\rightarrow\; I \otimes X, \; I \otimes Y, \; I \otimes Z, \; X \otimes I; X \otimes X; X \otimes Y; X \otimes Z; Y \otimes I; Y \otimes X; Y \otimes Y; Y \otimes Z; Z \otimes I; Z \otimes X; Z \otimes Y; Z \otimes Z\)\newline
    \textit{(16 equally likely two-qubit error outcomes, each with probability }p/15\textit{)} \\
    \hline
    \end{tabular}
\end{table}

\begin{table}[ht]
    \centering
    \caption{Uniform noise model used in all simulations. Each ideal gate (left column) is replaced by the corresponding noisy gate (right column). Noise channels are defined in Table~\ref{tab:noise-channels}.}
    \label{tab:ideal-noisy-gates}
    \renewcommand{\arraystretch}{1.2}
    \begin{tabular}{l l}
    \hline
    \textbf{Ideal Gate} & \textbf{Noisy Gate} \\
    \hline
    Idle       & \(\text{DEP1}(p)\) \\
    \(M_X\)    & \(\text{DEP1}(p) \cdot \text{MERR}_X(p)\) \\
    \(M_Z\)    & \(\text{DEP1}(p) \cdot \text{MERR}_Z(p)\) \\
    \hline
    \end{tabular}
\end{table}

Note that \(\text{MERR}_{X}(p)\) and \(\text{MERR}_{Z}(p)\) flip measurement outcomes of single-qubit \(X\)- or \(Z\)-type measurements with probability \(p\), while \(\text{MERR}_{XX}(p)\) and \(\text{MERR}_{ZZ}(p)\) do the same for two-qubit \(XX\)- or \(ZZ\)-type measurements.

\section{Correctable regions}\label{app:3d_plot}

\begin{figure*}
    \begin{tikzpicture}
        \node at (-4,3) {\includegraphics[width=.48\textwidth]{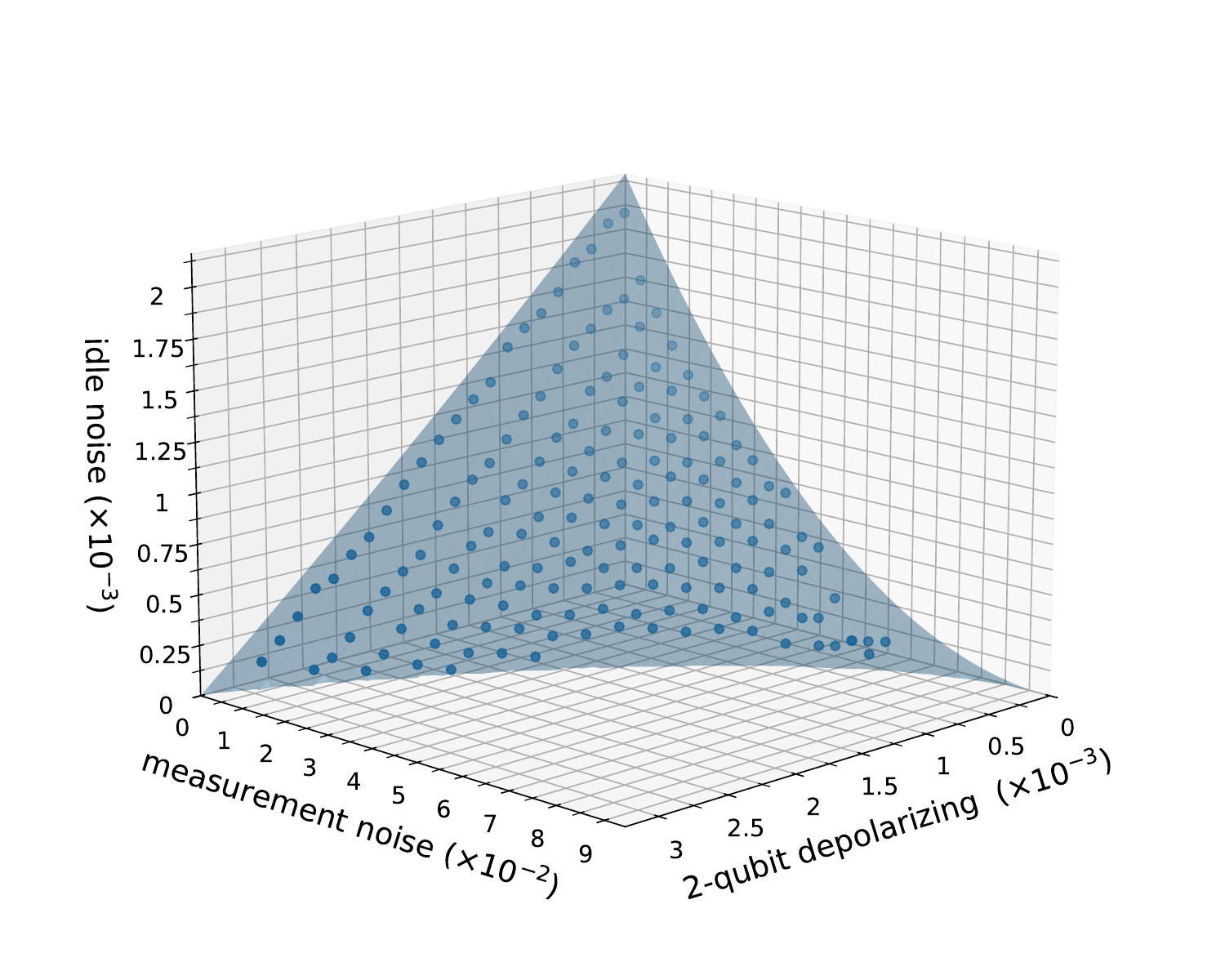}};
        \node at (4,3) {\includegraphics[width=.48\textwidth]{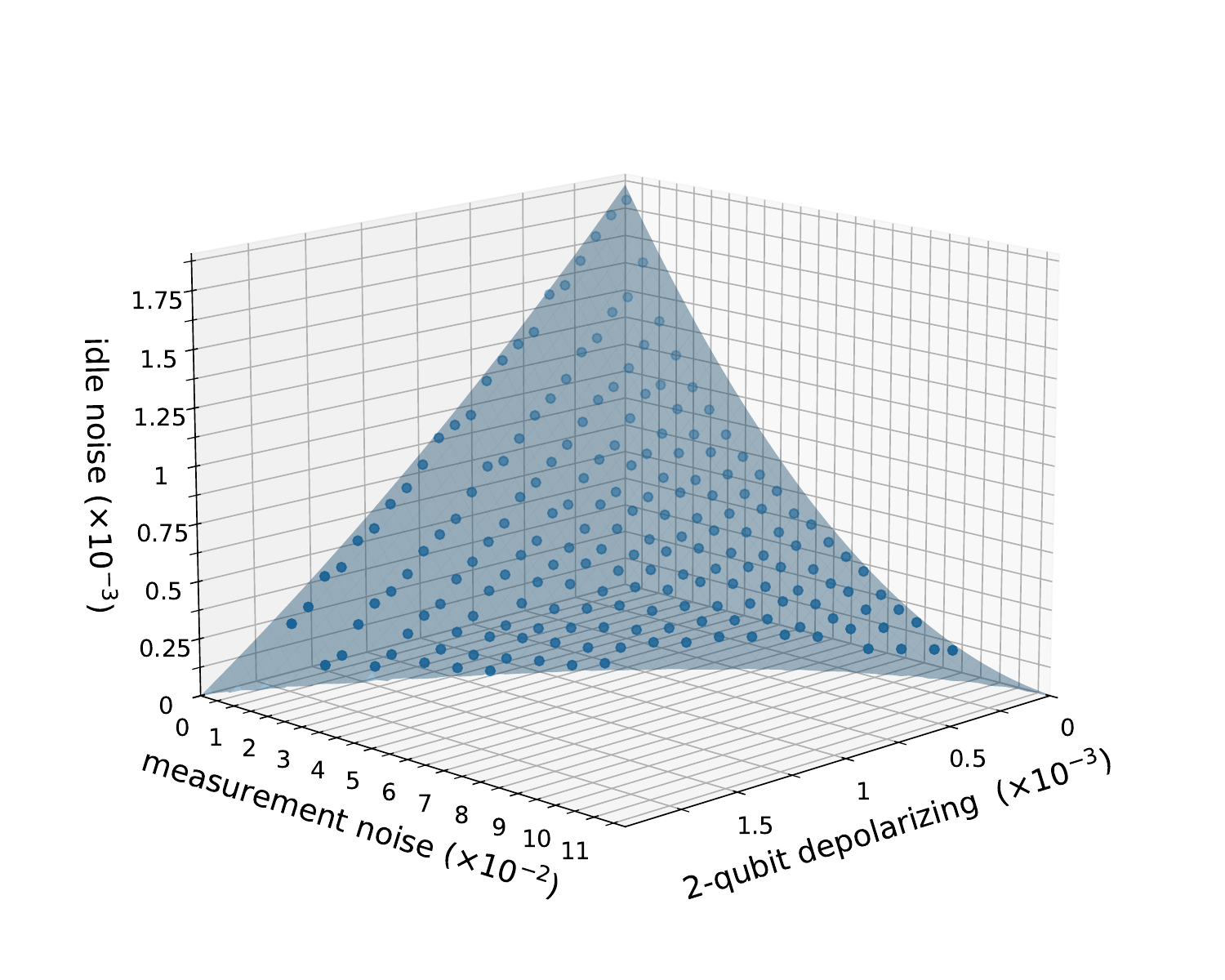}};
        \node at (-4,-4) {\includegraphics[width=.48\textwidth]{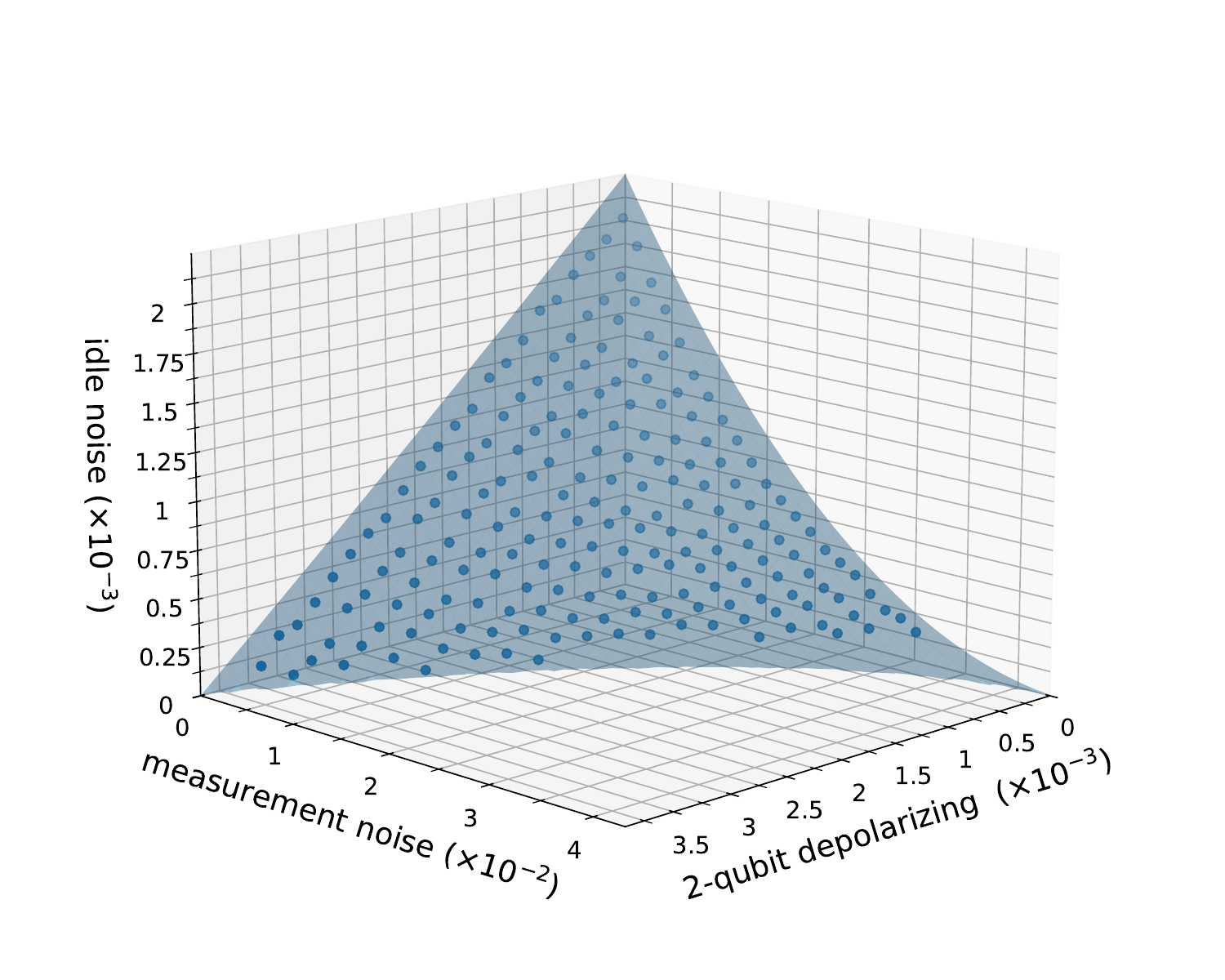}};
        \node at (4,-4) {\includegraphics[width=.48\textwidth]{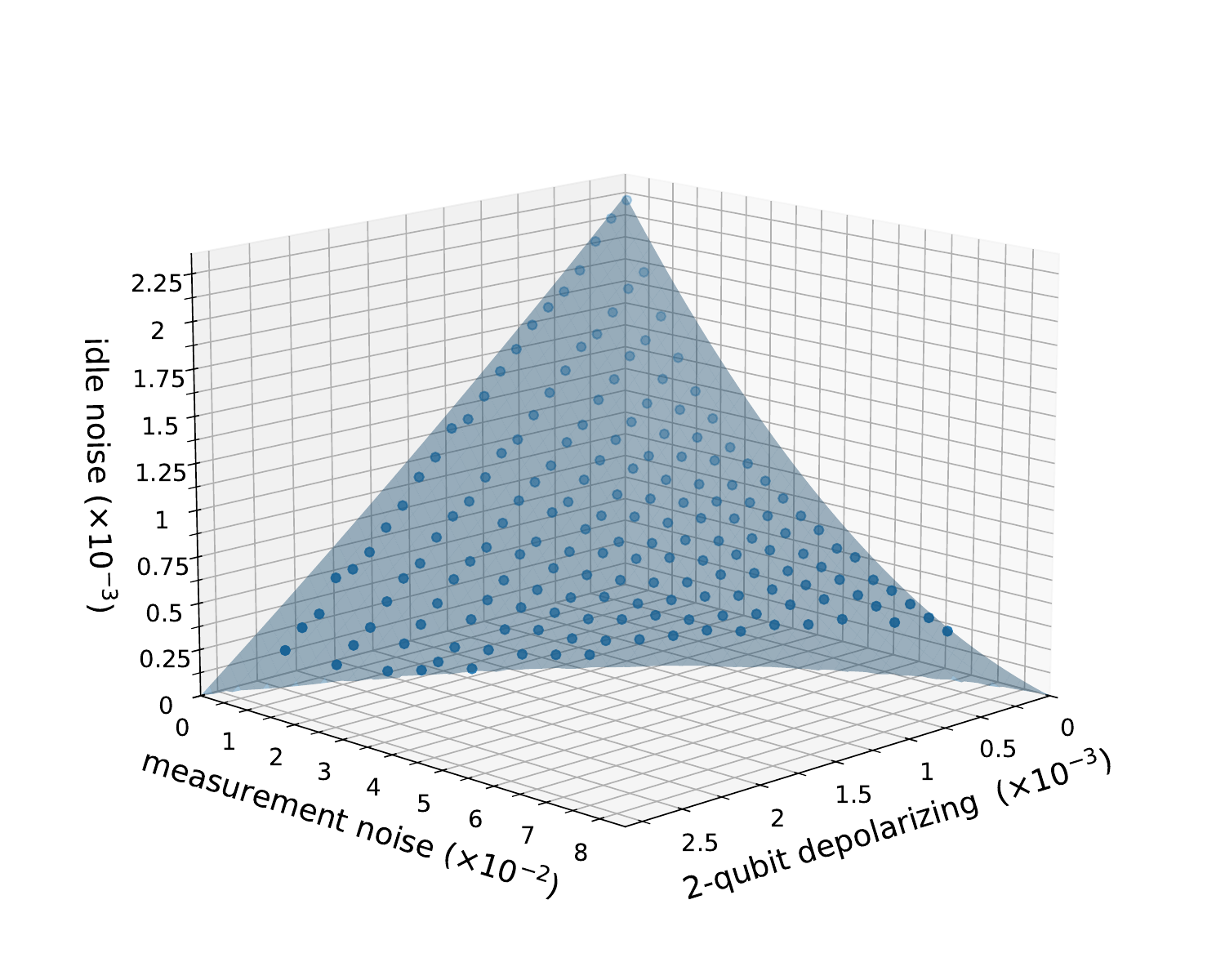}};
        \node at (-4,6) {\Large $X$ basis};
        \node at (4,6) {\Large $Z$ basis};
        \node [rotate=90] at (-8.5,3) {\Large no reset};
        \node [rotate=90] at (-8.5,-4) {\Large reset};
    \end{tikzpicture}
    
    \caption{
        Estimated threshold surfaces for the dynamic compass code (rightmost schedule in \cref{fig:schedules}) under a range of noise models. The top row shows results without reset, while the bottom row shows results with reset. In each row, the left panel corresponds to the $X$ basis and the right panel to the $Z$ basis. The single-qubit depolarizing error rate is fixed at $0.02\%$, consistent with the experimental value reported in Ref.~\cite{harper2025characterising}, while the remaining noise parameters are varied. The threshold surface (shown in blue) is obtained via quadratic fitting. The region below each surface defines the correctable regime, where the logical error rate decreases exponentially with increasing code distance. Above the surface, error correction fails due to excessive noise.
        }
    \label{fig:3d}
\end{figure*}
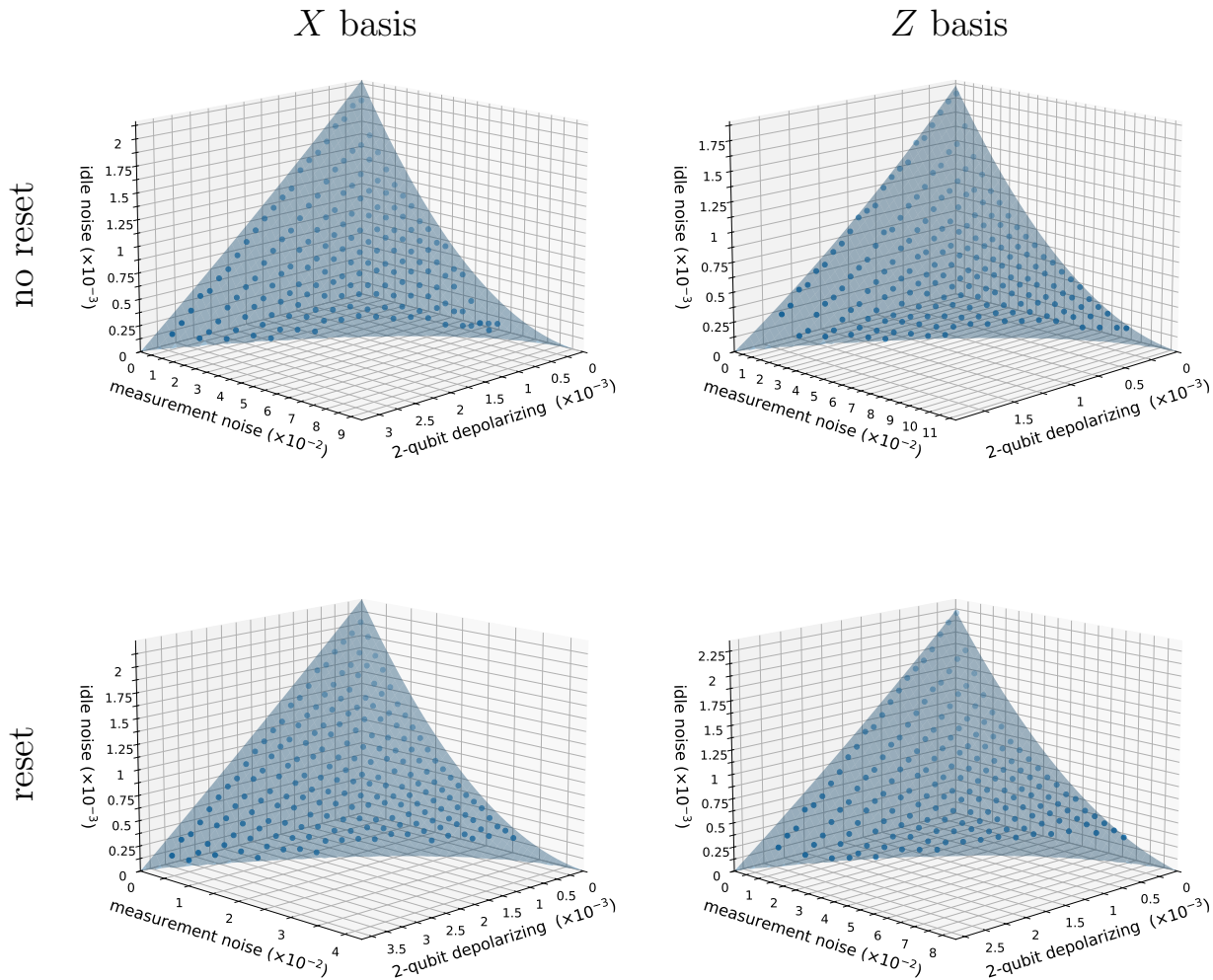

We fix a schedule (schedule C from \cref{fig:schedules}) and simulate its performance under a wide variety of asymmetric error models in order to map out the boundary between correctable and uncorrectable regions for this code. The results are presented in \cref{fig:3d} where we show three-dimensional threshold surfaces bounding correctable regions for $X$ and $Z$ errors in both reset and no-reset circuit situations. Details of the threshold estimation process are provided in~\cref{appdx:approx-method}. 

In both implementations, the code is significantly more resilient to measurement noise than to $2$-qubit depolarizing or idling noise, which is likely due in part to the design of the circuits we use. We also observe that the no-reset implementation tolerates substantially higher measurement noise than the reset implementation, by approximately a factor of two in the $X$ basis and by approximately a factor of $1.3$ in the $Z$ basis. This is consistent with the fact that reset operations introduce additional measurements. In heavy-hex devices, such resets are typically implemented via measurements followed by conditional bit flips, and were identified as a major source of noise in previous work~\cite{harper2025characterising}. We do not have a clear understand of the reason for the difference between the two bases but is likely due in part to the more complex ordering of $Z$ measurements, as well as that the $Z$ detectors have a greater temporal extent than the $X$ detectors.

We also observe a basis-dependent asymmetry in the performance against $2$-qubit depolarising and measurement noise. In particular, performance against $2$-qubit depolarising errors is better in the $X$ basis than in $Z$ in both reset and no-reset case which can be attributed to the fact that $X$ measurement qubits are involved in less two-qubit gates than $Z$ measurement qubits. On the other hand, measurement errors have more significant effect when decoding in the $X$ basis. One possible explanation for this is that measurements contributing to $Z$ detectors happen less frequently, and so in regimes where measurement errors are the dominant noise source it is actually favourable to measure stabilisers less frequently. Interestingly, performance against idling noise in the $X$ basis seems independent of resets, while this is not the case in the $Z$ basis.

\begin{figure*}
    \centering
    \begin{tikzpicture}
        \node (A) at (-4,4) {\includegraphics[width=0.4\linewidth]{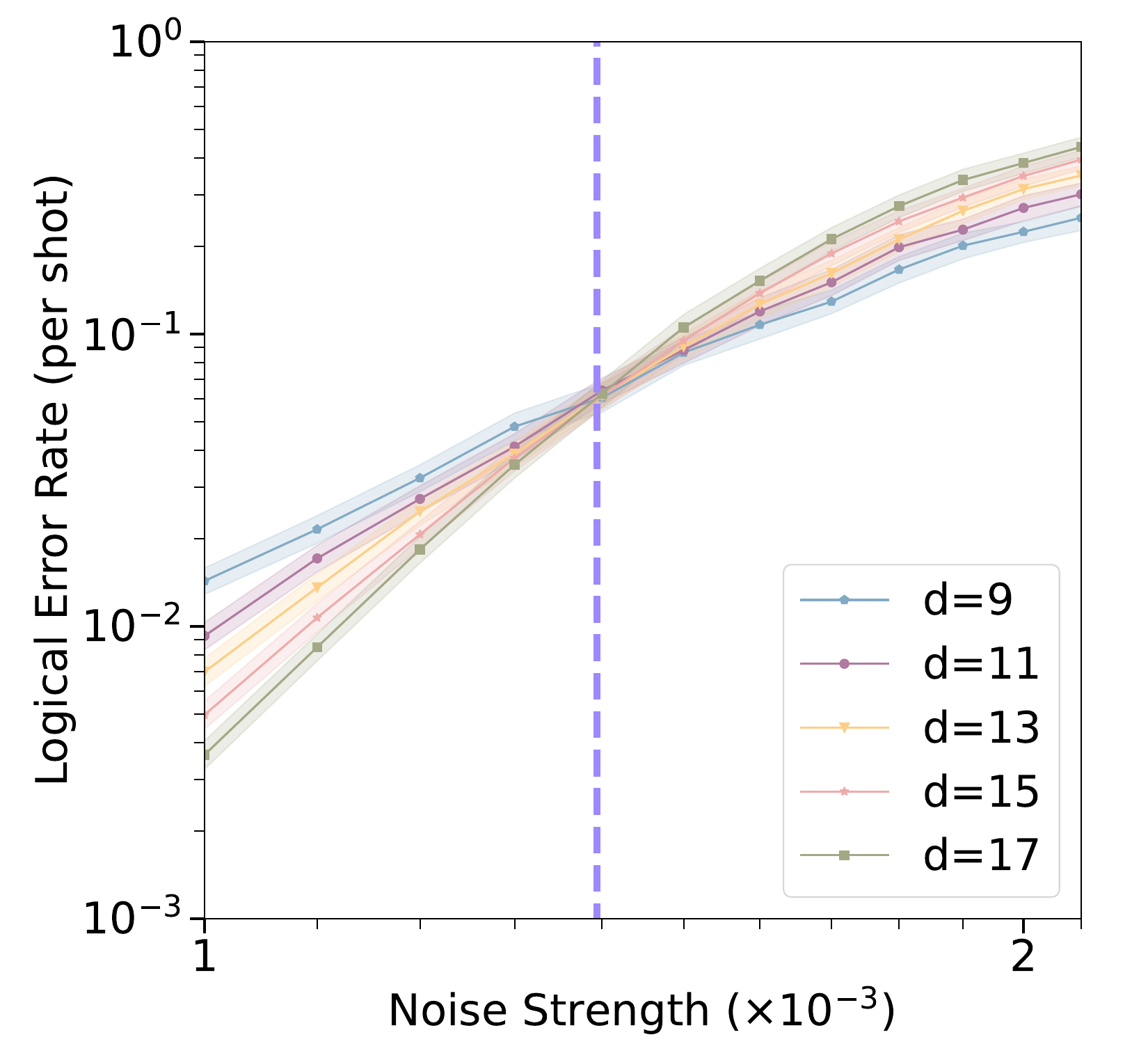}};
        \node (B) at (4,4) {\includegraphics[width=0.4\linewidth]{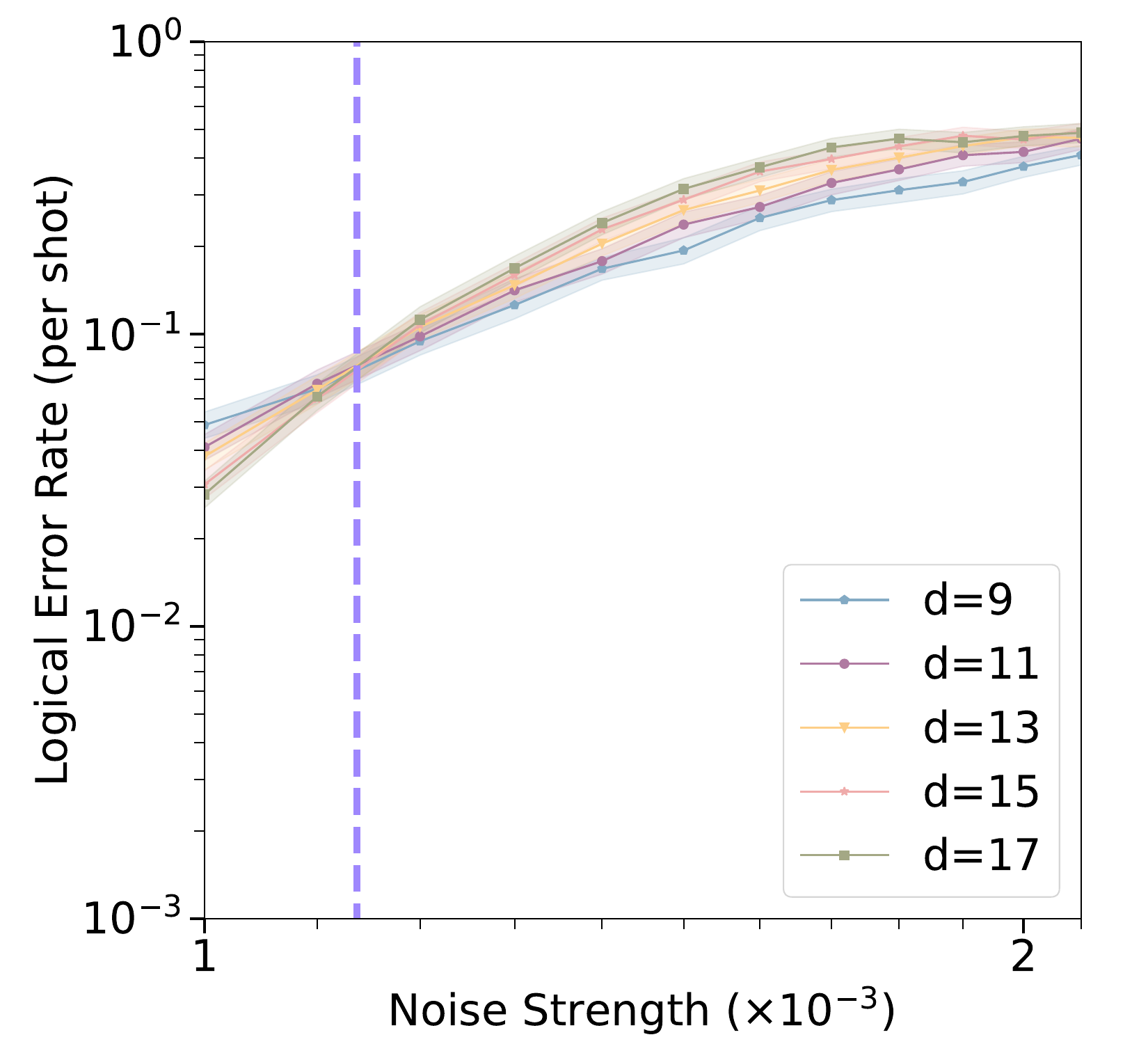}};
        \node (C) at (-4,-4) {\includegraphics[width=0.4\linewidth]{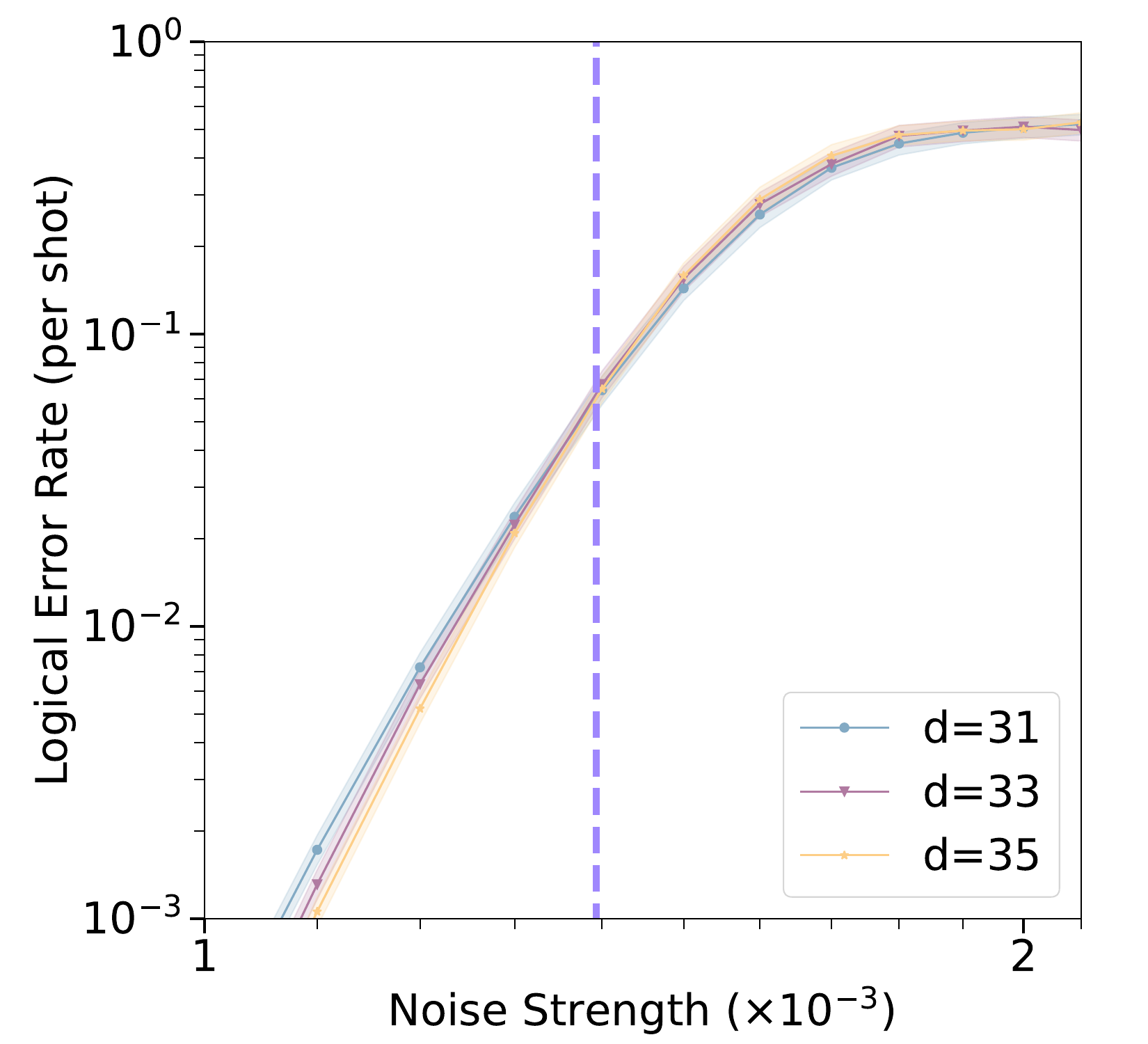}};
        \node (D) at (4,-4) {\includegraphics[width=0.4\linewidth]{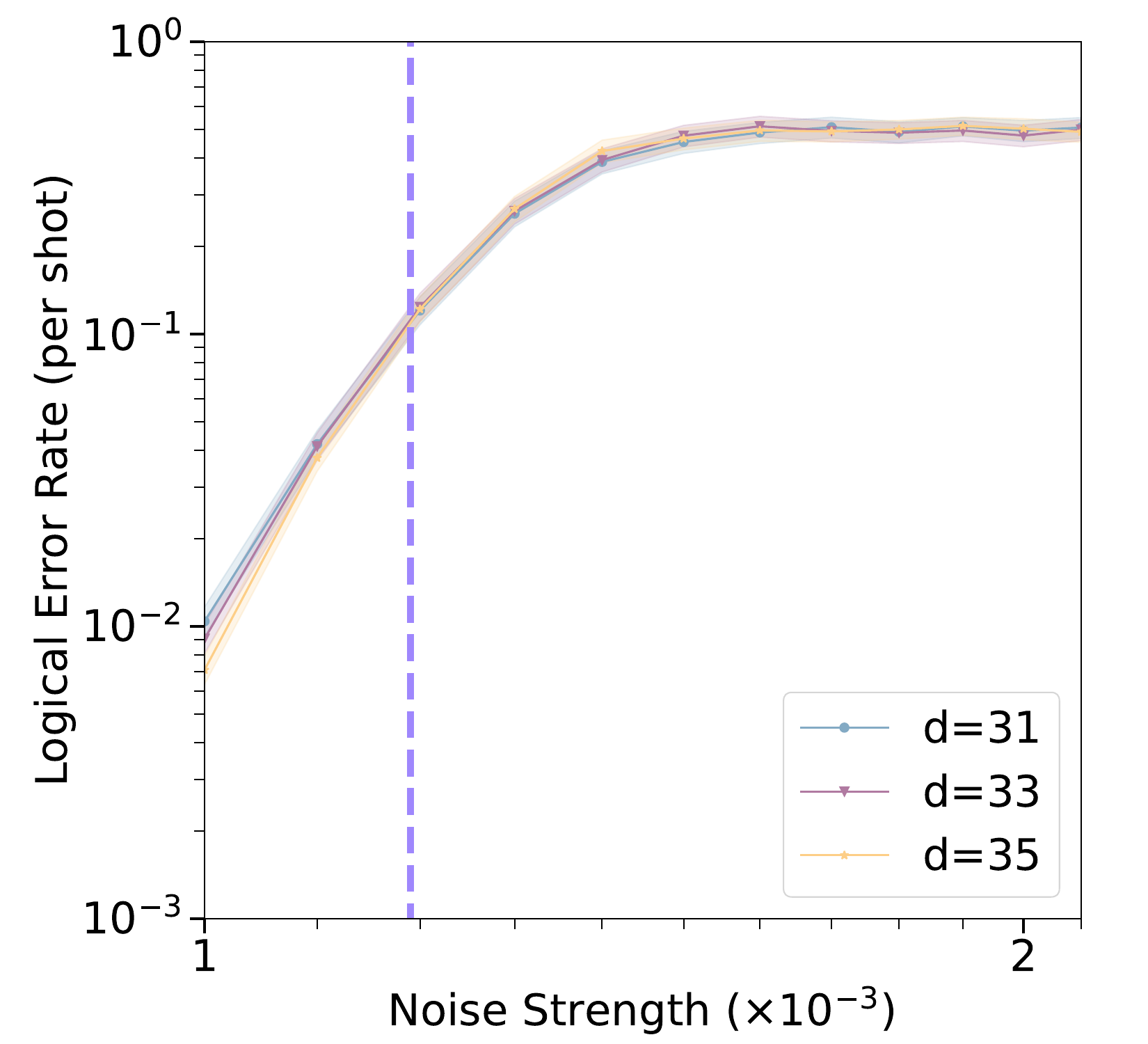}};
        \node at (A.north west) {(a)};
        \node at (B.north west) {(b)};
        \node at (C.north west) {(c)};
        \node at (D.north west) {(d)};
    \end{tikzpicture}
    \caption{Comparison of threshold estimates in ((a) and (c)) $X$ and ((b) and (d)) $Z$  basis obtained from small code distances $(d = 9, 11, 13, 15, 17)$ and larger code distances $(d = 31, 33, 35)$. Threshold points are indicated by dashed lines.}
    \label{fig:threshold-small-vs-large}
\end{figure*}
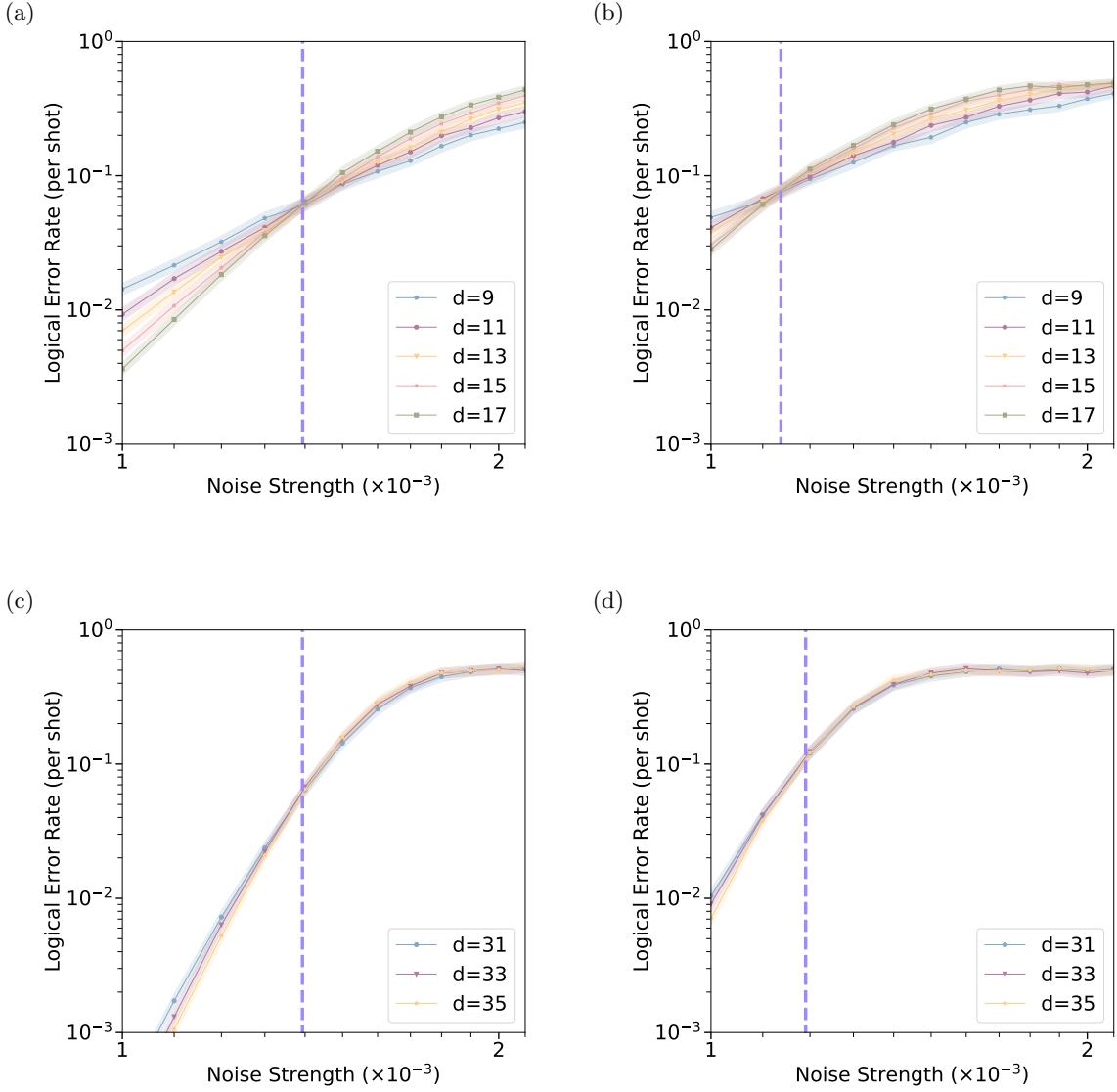

\section{Approximate threshold estimation}
\label{appdx:approx-method}

Accurate threshold estimation methods \cite{brown2016gaugecolor} require dense sampling in the area of the threshold. In the context of constructing the three-dimensional threshold surface in \cref{fig:3d}, this would require a large number of simulations over a wide range of noise parameters.

To make this computation tractable, we instead use an approximate method to estimate the threshold. In \cref{fig:3d}, each threshold point is obtained by fixing the two-qubit depolarizing noise and measurement noise, and sweeping the idling noise. For each such configuration, we obtain logical error-rate curves as a function of the swept error rate $p$ for multiple code distances.

We estimate the threshold by identifying crossings between these curves. Specifically, we scan adjacent intervals $[p_k, p_{k+1}]$ and detect crossings between pairs of code distances by checking for a change in the ordering of their logical error rates between the two endpoints.

When a crossing is detected, its location is estimated within the interval by interpolation. The interpolation is performed either in linear space or in log-log space, depending on how the physical error rates are sampled. When the sweep over $p$ is approximately uniform in logarithmic scale, we perform interpolation in log-log space; otherwise, we use linear interpolation.

Each crossing provides a candidate estimate for the threshold. We collect all such candidates across the parameter range and identify the densest cluster of crossing points, either in linear space or logarithmic space. A threshold is reported only when this cluster contains a large fraction (typically $\gtrsim 70\%$) of the candidates. The threshold value is then taken as the median of the points within this cluster.

To avoid including crossing from highly noisy regimes, where the logical error rate fluctuate around $0.5$, we exclude intervals where all logical error rates exceed a fixed cutoff. We use a cutoff of $0.4$ to exclude intervals where the logical error rate is close to saturation. In this regime, the logical error rate approaches $0.5$ and crossings are dominated by statistical fluctuations.

If no consistent cluster of crossings is found, we conclude that no approximate threshold is observed within the sampled range.

The threshold results in \cref{fig:3d} are obtained using the estimation procedure described above, based on simulations with code distances $d = 9, 11, 13, 15, 17$. To assess the validity of using only small code distances, we compare threshold estimates obtained from small code distances with those obtained from larger distances in \cref{fig:threshold-small-vs-large}. We observe that the estimated thresholds from small and large code distances are in close agreement, as indicated by the dashed lines.

\section{Stability Error Suppression}\label{app:stab_error_supp}

\begin{figure}
    \centering
    \includegraphics[width=0.9\linewidth]{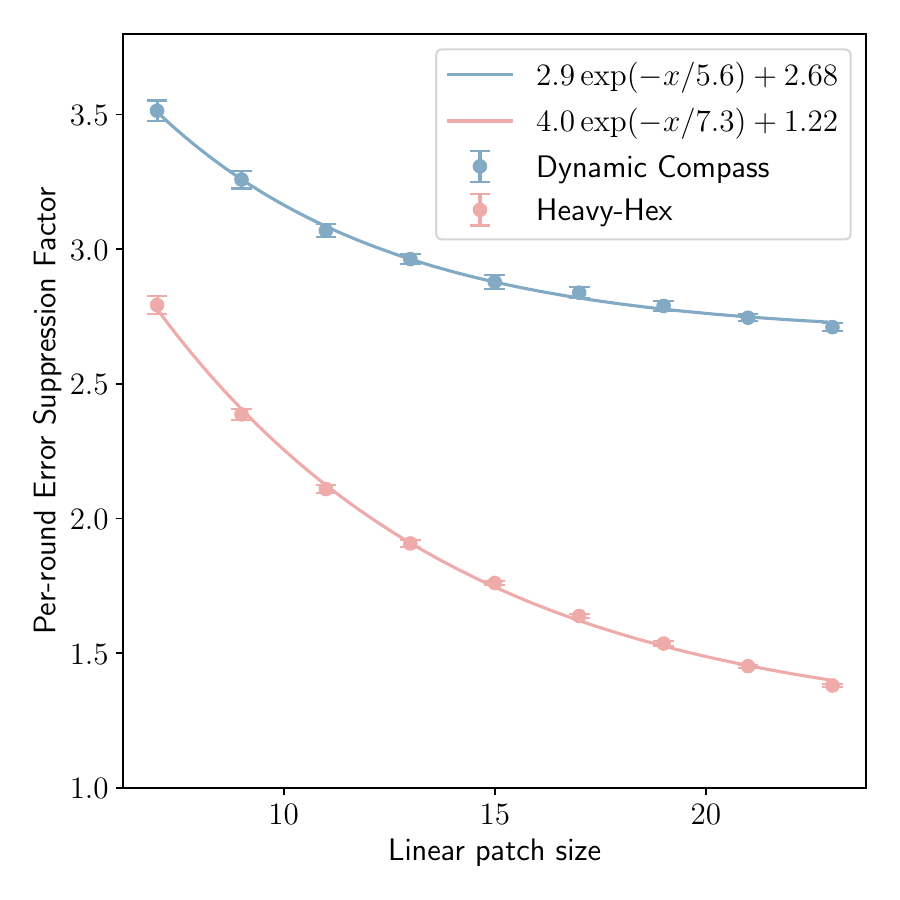}
    \caption{Error suppression factors and exponential decay fits for the dynamic compass and heavy-hex codes. Error bars indicate standard deviations. Fits are performed to the function $A \exp(-x/\tau) + B$, with obtained values for the dynamic compass code: $A = 2.9 \pm 0.2$, $\tau = 5.6 \pm 0.3$, $B = 2.68 \pm 0.02$. For the heavy-hex code we find: $A = 4.0 \pm 0.1$, $\tau = 7.3\pm 0.4$, $B = 1.22\pm 0.03$.}
    \label{fig:Stability_exp_decay}
\end{figure}

Here we display evidence that the per-round logical error suppression factor in stability experiments exponentially approaches a limiting value $>1$ (in the case of the dynamic code) as we increase the patch size. We focus on a value of $p$ well-below the stability threshold point. In \cref{fig:Stability_exp_decay}, we display data for the error suppression factor $s(d,p)$ as a function of the patch size, $d$, with fixed noise strength $p = 4\times 10^{-4}$. We fit the data to exponential decay curves, with good agreement, indicating that the dynamic compass code's suppression factor asymptotes to $\sim 2.68 \pm 0.02$, for this value of $p$. The heavy-hex's suppression factor asymptotes to $\sim 1.22 \pm 0.03$, although we do not expect this to be above $1$ in the long-time limit, and hence we may be limited by insufficient data.

\end{document}

%% file: main.bbl
%